\documentclass[10pt,twocolumn]{article}
\usepackage[top=1.9cm, bottom=1.9cm, left=1.9cm, right=1.9cm]{geometry}
\usepackage{times}  %
\usepackage[sort&compress,numbers]{natbib}  %
\usepackage[hyphens]{url}
\usepackage{graphicx}  %
\usepackage[keeplastbox]{flushend}
\usepackage[hyphens]{url}  %
\usepackage{graphicx} %
\usepackage{tikzsymbols}

\newcommand{\descr}[1]{\smallskip\noindent\textbf{#1}}

\usepackage{sectsty}
\sectionfont{\bfseries\Large\raggedright}

\usepackage{url}                                %
\usepackage{array,multirow,graphicx,adjustbox}  %
\usepackage{booktabs}                           %
\usepackage[utf8]{inputenc}
\usepackage[ruled,algosection,noend,linesnumbered]{algorithm2e}
\usepackage{float}
\usepackage{paralist}
\usepackage{amsmath}
\usepackage{amsfonts}
\usepackage{xcolor}
\usepackage{csquotes} 
\usepackage{tabularx}
\usepackage{makecell}
\usepackage{pbox}
\usepackage[hang,flushmargin]{footmisc}
\usepackage{flushend}
\usepackage{xspace}
\usepackage{multicol, lipsum}
\usepackage[T1]{fontenc}

\usepackage{xcolor}
\usepackage{booktabs}
\usepackage{graphicx}
\usepackage{paralist}
\usepackage[small,bf]{caption}
\usepackage{subfigure}

\newcommand{\dspol}{{{\selectfont /pol/}}\xspace}
\newcommand{\greatawakening}{{{\selectfont /v/GreatAwakening}}\xspace}
\newcommand{\news}{{{\selectfont /v/news}}\xspace}
\newcommand{\qresearch}{{{\selectfont /qresearch/}}\xspace}

\let\oldbibliography\thebibliography
\renewcommand{\thebibliography}[1]{%
  \oldbibliography{#1}%
  \setlength{\itemsep}{2pt}%
}

\usepackage[hang,flushmargin]{footmisc}

\usepackage[compact]{titlesec}
\titlespacing*{\section}{0pt}{*3}{3pt}
\titlespacing*{\subsection}{0pt}{*2}{2pt}
\usepackage{xspace}

\makeatletter
\def\url@leostyle{%
  \@ifundefined{selectfont}{\def\UrlFont{}}%
  {\def\UrlFont{}}%
}
\makeatother
\usepackage[hyphenbreaks]{breakurl}

\usepackage[bookmarks=true, bookmarksnumbered=true, colorlinks=true, linkcolor=linkcol, citecolor=citecol, urlcolor=urlcol, hypertexnames=true]{hyperref}

\definecolor{darkgreen}{RGB}{0, 100, 0}
\definecolor{linkcol}{rgb}{0.3,0,0}
\definecolor{citecol}{rgb}{0.3,0,0}
\definecolor{urlcol}{rgb}{0.3,0,0}

\makeatletter
\def\url@leostyle{%
  \@ifundefined{selectfont}{\def\UrlFont{\small}}%
  {\def\UrlFont{}}%
}
\makeatother
\begin{document}

\title{\bf The Gospel According to Q: Understanding the QAnon Conspiracy from the Perspective of Canonical Information\thanks{Published in the Proceedings of the 16th International AAAI Conference on Web and Social Media (ICWSM 2022). Please cite accordingly.}} 

\author{Antonis Papasavva$^{\ddag}$, Max Aliapoulios$^{\dag}$, Cameron Ballard$^\dag$, Emiliano De Cristofaro$^\ddag$,\\ 
Gianluca Stringhini$^\diamond$, Savvas Zannettou$^\circ$, and Jeremy Blackburn$^\mp$\\[0.5ex]
\normalsize $^\ddag$University College London, $^{\dag}$New York University, $^\diamond$Boston University,\\[-0.5ex]
\normalsize$^\circ$Delft University of Technology, $^\mp$Binghamton University\\
\normalsize antonis.papasavva@ucl.ac.uk, maliapoulios@nyu.edu, clb478@nyu.edu, e.decristofaro@ucl.ac.uk,\\[-0.5ex]
\normalsize gian@bu.edu, s.zannettou@tudelft.nl, jblackbu@binghamton.edu}

\date{}

\maketitle

\begin{abstract}

The QAnon conspiracy theory claims that a cabal of (literally) blood-thirsty politicians and media personalities are engaged in a war to destroy society.
By interpreting cryptic ``drops'' of information from an anonymous insider calling themself Q, adherents of the conspiracy theory believe that Donald Trump is leading them in an active fight against this cabal.
QAnon has been covered extensively by the media, as its adherents have been involved in multiple violent acts, including the January 6th, 2021 seditious storming of the US Capitol building.
Nevertheless, we still have relatively little understanding of how the theory evolved and spread on the Web, and the role played in that by multiple platforms.

To address this gap, we study QAnon from the perspective of ``Q'' themself.
We build a dataset of 4,949 canonical Q drops collected from six ``aggregation sites,'' which curate and archive them from their original posting to anonymous and ephemeral image boards.
We expose that these sites have a relatively low (overall) agreement, and thus at least some Q drops should probably be considered apocryphal. 
We then analyze the Q drops' contents to identify topics of discussion and find statistically significant indications that drops were not authored by a single individual.
Finally, we look at how posts on Reddit are used to disseminate Q drops to wider audiences.
We find that dissemination was (initially) limited to a few sub-communities and that, while heavy-handed moderation decisions have reduced the overall issue, the ``gospel'' of Q persists on the Web.
\end{abstract}

\section{Introduction}
While ubiquitous social media has helped foster new relationships and disseminate information, not everything is beneficial to society.
Over the past decade, a few conspiracy theories have emerged, often blaming secret organizations, governments, or cabals for world-changing events~\cite{911}, %
E.g., conspiracy theorists claim that Bill Gates created the COVID-19 pandemic to implant microchips in people via the worldwide administration of a vaccine~\cite{billgates}.
Some of these theories can threaten democracy itself~\cite{sternisko2020dark,schabes2020birtherism}; e.g., \emph{Pizzagate} emerged during the 2016 US Presidential elections and claimed that Hillary Clinton was involved in a pedophile ring~\cite{pizzagate}.

A specific example of the negative consequences social media can have is the QAnon conspiracy theory.
It originated on the Politically Incorrect Board (\dspol) of the anonymous imageboard 4chan via a series of posts from a user going by the nickname Q.
Claiming to be a US government official, Q described a vast conspiracy of actors who have infiltrated the US and other governments worldwide waging war against freedom, and another set of actors, led by Donald Trump, actively fighting back~\cite{whatisQAnon1}.
Since its inception in 2017, it has grown to encompass numerous existing conspiracies, including Pizzagate.%

QAnon has long ceased to be an inconsequential conspiracy theory confined to the Internet's dark corners.
The events of January 6th, 2021, when a pro-Trump mob rushed the US Capitol, demonstrate how deeply entrenched QAnon is in violent calls to far-right extremist actions~\cite{capitolriot2}.
In the aftermath of the insurrection, it became clear that many of the people involved were QAnon followers, including law enforcement officers, former military, and Internet personalities~\cite{militaryINcapitolriot}.
Even before, QAnon supporters had been linked to various crimes, including an attempt to blow up a statue in Illinois, kidnapping children to ``save them from the pedophiles,'' etc.~\cite{QAnoncrimes}.

Overall, conspiracy theories can pose substantial risks to democratic societies, e.g., when used to benefit political agendas and interests~\cite{schabes2020birtherism}.
QAnon has proven this to great effect, as at least 25 US Congressional candidates with direct links to QAnon appeared on ballots during 2020 US House of Representatives elections~\cite{qanonCongressMembers}, and at least two elected US House Representatives publicly supported the movement~\cite{qanonCongressMembers2}.

Although having received ample media coverage, we still lack an understanding of how QAnon works, making it challenging to develop mitigation techniques for future conspiracies and directly address QAnon.
A primary challenge is directly related to QAnon's origin and evolution on imageboards like 4chan and 8chan/8kun.
Imageboards are ephemeral and anonymous, with the only method of persistent identification across posts being a fallible system known as \emph{tripcodes}.
Interestingly, QAnon adherents %
developed a set of sites that aggregate and ``authenticate'' messages posted by Q, known as \emph{Q drops}.
These Q drops are discussed on imageboards, collected on these aggregation sites for ease of access, and later discussed on other Web communities.

\descr{Problem Statement.}
In this paper, we set out to provide a broad understanding of the QAnon conspiracy theory through the lens of Q drops.
We study how and where the drops are cataloged and detect writing habit differences across tripcodes.
Going deeper into the conspiracy, we aim to elicit the main discussion topics of the conspiracy and whether or not these posts are toxic, threatening, and easy to understand.
Finally, we turn our attention to mainstream and alternative Web communities to shed light on how the conspiracy spread and evolved on other social networks.

\descr{Research Questions.} Our work is driven by the following research questions:
\begin{itemize}
\item[\bf RQ1] How does the canonicalization process of the QAnon conspiracy work? 
\item[\bf RQ2] What topics do canonical Q drops discuss, and what ideas does this content convey to adherents?
\item[\bf RQ3] How and where is the canonical Q content shared on social media?
\end{itemize}

\descr{Methodology.} We collect and analyze 30,320 Q drops (4,961 unique) from six aggregation sites, and the corresponding 4chan and 8chan/8kun threads that Q posted in, and 1.4M and 546K posts from Reddit and Voat.

To answer RQ1, we measure the agreement across all aggregation sites using Fleiss' kappa score~\cite{fleiss1971measuring} and calculate the set of overlapping drops across aggregation sites to build a canonical set.
In addition, we employ basic stylometric techniques to measure the similarity of posts across tripcodes.
For RQ2, we use word embeddings %
and Google's Perspective API, to analyze how different words are used in the Q drops, how they are interconnected, what are the various topics of interest, and how toxic and coherent is the content created by Q.
Finally, to answer RQ3, we study how aggregation links are mentioned on Reddit.

\descr{Main findings.}
Overall, we make the following findings:
\begin{itemize}
\item The six aggregation sites devoted to archiving Q drops have poor agreement scores between them. 
We detect significant differences regarding the writing habits of the five most-used tripcodes, which suggests there is no single canonical Q.
\item Q discusses, among other things, the ``usurpation'' of the government.
Q drops are also exceptionally incoherent, a likely explanation for the decoding/interpretation efforts of adherents. 
Although adherents have been involved in violence, Q drops are not particularly toxic or threatening.
This questions whether by themselves they may be considered high risk, at least by automated moderation tools; rather, toxicity and calls for violence stem from the interpretations of adherents and the actors with vested interests that weaponize it.
\item We find that the aggregation links were disseminated across Reddit from a handful of users.
Also, although Reddit banned QAnon-related subreddits, other subreddits, e.g., r/conspiracy, still share and discuss Q drops.
\end{itemize}

\section{Background and Related Work}
This section provides background information on the history and main beliefs of the QAnon movement and the Web communities that are part of our datasets. 
Finally, we review relevant previous work.

\subsection{QAnon}
On October 28, 2017, an anonymous user with the nickname ``Q'' posted a thread on 4chan's Politically Incorrect board (/pol/), titled ``Calm before the Storm,'' claiming to be a government insider with ``Q level'' security clearance.\footnote{The top-secret clearance for the US Department of Energy.}
Q claimed to have read documents proving the existence of a satan-worshiping pedophile cabal of highly influential and powerful people that secretly controls governments world wide~\cite{whatisQAnon2}.
Among other things, Q swears allegiance to an alleged noble crusade that Donald Trump is leading to bring this satanic cabal to justice.

\descr{Q Drops.} The posts that Q made on 4chan, and later 8chan/8kun, since 2017 are known as ``drops.''
QAnon followers devote themselves to decoding Q drops to understand and expose the actions of the ``deep state.''
The movement has since grown substantially on mainstream social networks like Facebook, Reddit, and Twitter. 
The conspiracy has even spread to countries other than the US, where QAnon adherents have staged protests~\cite{QAnonUK}.

\descr{QAnon Aggregation Sites.} Aggregation sites are platforms dedicated to providing a collective index of information about the conspiracy. 
They are created, developed, and funded by Q supporters to aggregate %
Q's drops and help others find information about the conspiracy.
The decision of which post is indeed a Q drop falls, to some extent, to the operators themselves.
Perhaps the most popular aggregation site is qmap.pub, which was shut down in September 2020 after an investigation led to identifying its owner and host~\cite{ondrak2020}.
Overall, aggregation sites are crucial data points for this study, as they provide insight into sources that report on and discuss the conspiracy.

\descr{QAnon on the Web.} 
As a conspiracy theory born out of the Web, it is not surprising that social networks have played a significant role in QAnon's evolution.
Active and fast-growing QAnon-related communities have emerged not only on fringe platforms but also on mainstream ones~\cite{menn2018}.
In fact, most of the latter have banned QAnon-linked groups and content---Reddit in 2018, and Twitter, Facebook, and YouTube in 2020~\cite{banQAnon}.
However, the deplatformed QAnon communities resurface on other fringe platforms like Voat, and discussion on 4chan and 8chan/8kun remains active~\cite{papasavva2021qoincidence}.%

\subsection{Web Communities}

\descr{Imageboards.} %
We collect and analyze data from 4chan and 8chan/8kun.
These are imageboards, anonymous and ephemeral social media where images are posted alongside text, organized in boards devoted to specific themes, e.g., sports, politics, etc.
Typically, users create a thread by posting an image and/or description, and others then can post on that thread with or without images.
We focus on 4chan and 8chan/8kun as the conspiracy started on 4chan's \dspol, before moving to 8chan in December 2018~\cite{Zeeuw2020normification}, which was shut down in August 2019~\cite{8chandown} and resurfaced in November 2019 as 8kun.
For simplicity, henceforth, we refer to both 8chan/8kun as 8kun.

Posts on imageboards are ephemeral (i.e., all posts and threads are deleted after some time) and, by default, anonymous, i.e., there are no user accounts.
Posts are displayed under the generic username \emph{Anonymous}, and users typically call each other as Anons---hence `QAnon' to refer to `Q.' 
However, users can choose a unique, linkable username for themselves using {\em ``tripcodes.''} 
Although 4chan and 8kun have different technical implementations, tripcodes are hashed passwords that allow users with the correct password to post under a username that makes them recognizable across threads~\cite{hine2017kek}.

\descr{Voat.} 
Voat was a Reddit-like news aggregation site, launched in April 2014 and shut down in December 2020. %
Voat often attracted users that had their hateful communities banned, e.g., r/CoonTown~\cite{redditTOvoat2}.
It also reportedly hosted QAnon-related communities banned from Reddit, like r/GreatAwakening~\cite{papasavva2021qoincidence}.
The Voat equivalent of a subreddit is called ``subverse.'' 

\subsection{Related Work}\label{sec:relatedwork}
Papasavva et al.~\cite{papasavva2021qoincidence} collect over 150K posts on QAnon-related Voat subverses, posted by 5K users in May--October 2020, finding that the QAnon community on Voat grew shortly after the Reddit bans.
They also show that conversations focus on world events, US politics, and Trump, while terms like QAnon and Q are closely related to Pizzagate.

Priniski et al.~\cite{priniski2021qanon} analyze 800K QAnon tweets from 2018, finding that the majority of users disseminates QAnon content, not creating it. 
Similarly, McQuillan et al.~\cite{mcquillan2020cultural} study QAnon on Twitter, finding that QAnon hashtags are associated with COVID-19; in fact, the Twitter QAnon community almost doubled in size between January and May 2020. 
Also, Darwish~\cite{darwish2018kavanaugh} analyze 23M tweets related to the US Supreme Court judge Brett Kavanaugh, finding that the hashtags \#QAnon and \#WWG1WGA\footnote{Where we go one we go all, a popular QAnon motto.} are among the top six hashtags in their dataset.
Chowdhury et al.~\cite{chowdhury2020twitter} collect 1M tweets from 2.4M suspended Twitter accounts, finding that politically motivated users consistently spread conspiracies, including QAnon.
Finally, Torres-Lugo et al.~\cite{torres2020manufacture} study ``follow trains'' (long lists of like-minded accounts that are mentioned for others to follow) on 5.5K Twitter accounts aiming to analyze political echo-chambers, and find that Republican users tweet QAnon-related hashtags often.

Aliapoulios et al.~\cite{aliapoulios2021parler} collect 120M posts from 2.1M users posted between 2018 and 2020 on Parler, an alternative social network that gained popularity after the 2020 US Elections and several conservative figures were banned from Twitter and Facebook. 
Among other things, they find that Parler's user base mainly consists of Trump supporters that are heavily discussing the QAnon conspiracy theory. 

Overall, this line of research focuses on single communities (such as, Twitter, Voat, Parler), %
whereas our work provides a multi-platform analysis of QAnon along several axes.
Furthermore, we do not only look at social network discussions, but at Q drops and aggregation sites as well.

OrphAnalytics~\cite{OrphAnalytics2020} analyze 4,952 Q drops collected from a single aggregation site (\url{qresear.ch}).
Using a (patented and undisclosed) unsupervised machine learning algorithm, they identify two individual signals, positing that drops were written by two different authors.
Our stylometric analysis (see Section~\ref{stylometry}) also suggests that the content written by the most used tripcodes originates from two different authors. 

Perhaps closer to our work is the study by De Zeeuw et al.~\cite{Zeeuw2020normification}, who collect QAnon-related data between October 2017 and November 2018 from 4chan's \dspol, 8chan's \qresearch, Reddit, Twitter, YouTube, and online press articles and comments.
They analyze the conspiracy theory's evolution from fringe communities to mainstream social networks and news.
They show that \dspol was the original board used by Q before it moved to \qresearch.
Around the same time, Reddit and YouTube users started mentioning the conspiracy increasingly often, while online press started covering it in-depth only after r/CBTS\_Stream got banned.

Our work differs from previous research in that we approach the problem from the perspective of Q drops themselves.
We are interested in understanding how Q drops are disseminated and canonicalized, comparing data across six aggregation sites and data from three social networks.
While other work has examined discussions and communities related to the conspiracy theory, there has been no systematic exploration of the ``source material,'' in terms of high-level topic and toxicity detection.
Furthermore, to the best of our knowledge, our multi-platform dataset is the largest and most complete to date.

\section{Datasets}\label{sec:datasets}
We now describe the data we collect and use in this work.

\descr{Q Drops.}
Using a custom crawler, we collect Q drops posted on six different QAnon aggregation sites between 2017 and 2020.
We find one of the most known aggregation sites (namely, qmap.pub) from a fact-checking site article~\cite{ondrak2020}; then, we extend to other aggregation sites linked there (some even share an open-source codebase.\footnote{The open-source code of the aggregation sites was published on GitHub but is currently down: \url{https://github.com/QAlerts/Sqraper}})
Therefore we argue that by following these hyperlinks, observing forks of boilerplate aggregation site codebases, and performing other open-source research, we enumerated the most popular mainstream aggregation sites at the time of writing.

Table~\ref{table:aggregation_sites} reports the number of Q drops per aggregation site.
Note that a drop is considered unique by its post ID and the specific board that it is posted on. 

  \begin{table}[t!]
    \centering
    \small
    \begin{tabular}{@{}lr@{}}
      \toprule
      \textbf{Aggregation Site} & \textbf{\#Drops} \\ \midrule
      qagg       &        4,954 \\
      qalerts    &        4,953 \\
      operationq &        4,953 \\
      qanon.news &        4,952 \\
      qanon.pub  &        4,854 \\
      qmap.pub       &    4,650 \\ \midrule
      \textbf{Total (unique)}            & 4,961 \\
      \bottomrule
      \end{tabular}
   \caption{Counts of collected drops across aggregation sites.}%
   \label{table:aggregation_sites}
   \end{table}

\descr{4chan and 8kun.}
Following the same methodology of Hine et al.~\cite{hine2017kek}, we collect threads and posts from 4chan and 8kun.
We focus on eight distinct boards that the aggregation sites note as containing posts from Q.
For each board, we collect all threads and posts made between June 2016 to November 2020.
Table~\ref{table:board_count} reports the number of threads and posts we collect for each board.

\begin{table}[t!]
\centering
\small
\begin{tabular}{@{}llrr@{}}
  \toprule
  \textbf{Board} & \textbf{Site} & \textbf{\#Posts} & \textbf{\#Threads} \\ \midrule
  pol            & 4chan         & 141,722,957      & 3,297,289          \\
  qresearch      & 8kun    & 10,661,799       & 16,729             \\
  pol            & 8kun    & 3,931,616        & 47,680             \\
  cbts           & 8kun    & 163,745          & 453                \\
  thestorm       & 8kun    & 35,828           & 124                \\
  patriotsfight  & 8kun    & 248              & 3                  \\
  projectdcomms  & 8kun    & 11               & 1                  \\
  greatawakening & 8kun    & 9                & 2                  \\ \bottomrule
  \end{tabular}
\caption{Post and thread count across all QAnon related boards.}
\label{table:board_count}
\end{table}

There are some gaps in our 4chan and 8kun datasets;
this is primarily due to infrastructure failures (recall that these platforms are ephemeral) and periods of sporadic availability when 8chan rebranded as 8kun.
We thus use data archived on \url{archive.org} to backfill as many gaps as possible.
Specifically, we collect 435,668 posts and 1,909 threads from \url{archive.org}, using the domain and thread IDs from 4,961 unique Q drop links on the aggregation sites.
Of the 4,961 drops, our crawlers retrieve 4,415 (88.99\%).
The missing 546 drops are likely due to crawling issues too, and we retrieve 99 of them from \url{archive.org}.

Finally, from the 1,936 total (unique) threads that aggregation sites claim drops were posted in, our crawlers retrieve 1,858 (95.97\%);
using the data from \url{archive.org}, we collect 67 of the missing threads.
Note that Table~\ref{table:board_count} includes the number of posts/threads obtained from \url{archive.org}.

\descr{Reddit.}
Reddit was one of the first mainstream social network to host and ban QAnon discussions~\cite{papasavva2021qoincidence}. 
Therefore, we detect and collect data from subreddits that promoted conspiracy material towards understanding, among other things, how the conspiracy spread to mainstream networks.
We start from all the data collected by Pushshift~\cite{baumgartner2020pushshift} between November 2017 and April 2020, extracting the 6,344 comments and 712 posts that contain a direct link to a drop or an aggregation site. 
We complement our Reddit dataset with all posts made on QAnon-related subreddits.
To find QAnon-related subreddits, we search the Pushshift archive for subreddits with names similar to the ones reported by previous work~\cite{papasavva2021qoincidence} and online press related to QAnon~\cite{wyrich2018,redditTOvoat}.
As discussed in Section~\ref{sec:drop_spread}, we use links to aggregation sites as a way of measuring the conspiracy spread on Reddit.
Overall, we collect 121,956 posts and 1,304,523 comments shared on Reddit between November 2017 and April 2020 (see Table~\ref{table:subreddit_collected_data}).

\begin{table}[t]
  \centering
  \small
  \begin{tabular}{l r r}
    \toprule
    \textbf{Subreddit}    & \textbf{\#Posts}            & \textbf{\#Comments}           \\ \midrule
    greatawakening        & 79,952                      & 926,676                       \\
    CBTS\_Stream          & 30,176                      & 267,744                       \\
    Qult\_Headquarters    & 7,465                        & 101,776                       \\
    The\_GreatAwakening   & 1,479                       & 0                             \\
    eyethespyzone         & 1,292                       & 1,398                         \\
    AFTERTHESTQRM         & 648                         & 1,544                         \\
    TheGreatAwakening     & 343                         & 15                            \\
    BiblicalQ             & 274                         & 551                           \\
    WakeAmericaGreatAgain & 124                         & 135                           \\
    QAnon                 & 122                         & 339                           \\
    QProofs               & 76                          & 254                           \\
    CalmBeforeTheStorm    & 5                           & 3                             \\ \midrule
    Aggregation filtering    & 714                           & 6,344                             \\ \midrule
    \textbf{Total (unique)}        & 121,956 & 1,304,523 \\ \bottomrule
    \end{tabular}
  \caption{Comments and posts collected from Reddit. All posts and comments crawled using aggregation link filtering are grouped.}
  \label{table:subreddit_collected_data}
 \end{table}

\descr{Voat.} We use the methodology of Papasavva et al.~\cite{papasavva2021qoincidence} to collect submissions and comments from the \greatawakening and \news subverses, between May 28 and December 10, 2020.
Overall, we collect 21,668 submissions and 196,673 comments from \greatawakening, and 141,177 submissions and 186,595 comments from \news.
Similar to Papasavva et al.~\cite{papasavva2021qoincidence}, we use the data from \news as a baseline dataset and the data from \greatawakening as standard QAnon discussion dataset for later analyses in Section~\ref{sec:toxicity}.

\descr{Ethical considerations.}
We only collect publicly available data, following ethical guidelines that are common in the computational social science community~\cite{rivers2014ethical} as well as ethical principles constituting the de-facto standard in the USA~\cite{dittrich2012menlo}.
Also, note that we do not attempt to identify users or link profiles across platforms. 
Moreover, the collection of data analyzed in this study does not violate any of the social networks' Terms of Service.

\section{Canonicalization of QAnon}
We now compare the Q drops collected from six aggregation sites; we shed light on which drops these sites consider canonical and the agreement between them.
Then, we analyze Q's posts using different tripcodes stylometrically.

\subsection{Is There a Canonical Set of Q Drops?}\label{stylometry}
The first question we set out to answer is to what degree different aggregation sites agree on what constitutes a canonical Q drop.
This is important since aggregation sites provide an archival system for ephemeral data; also, over time, Q migrated across 3 imageboards. %
Even assuming a perfectly secure system, which tripcodes are not, this presents an opportunity for apocryphal drops to be introduced. 

Measuring discrepancies in Q drops and what could be considered canonical gives insight into the shared knowledge which fuels the conspiracy theory.
Also, since aggregations sites are the ``bible'' of the conspiracy, collecting and analyzing information these sites curate is crucial to understand the conspiracy at large.
This is because most QAnon adherents digest conspiracy related content from aggregation sites instead of the source itself~\cite{ondrak2020}.
Thus, we analyze the agreement of the content across 6 different aggregation sites to explore one of the gateways through which the conspiracy spread to the mainstream.

\descr{Agreement between aggregation sites.}
We use a standard statistical measure of agreement quality, the Fleiss' kappa score~\cite{fleiss1971measuring}, which measures the agreement between multiple annotators on a classification task.
For our purposes, we treat each aggregation site as an annotator classifying whether or not a given Q drop (uniquely identified by the post ID and the board it appeared on) is canonical.

When computed across all Q drops in our dataset, we find \emph{poor} agreement ($\kappa < 0$) as per~\cite{koch1977}.
One of the major reasons for this is that qmap.pub was shut down in September 2020 and thus did not archive the several hundred Q drops that occurred later in 2020.
When we remove qmap.pub from the dataset, the Fleiss score increases ($\kappa = 0.24$).
While this is considered \emph{fair} agreement, there is enough discrepancy to warrant a deeper look.

\begin{figure}[t]
  \centering
  \includegraphics[width=0.9\columnwidth]{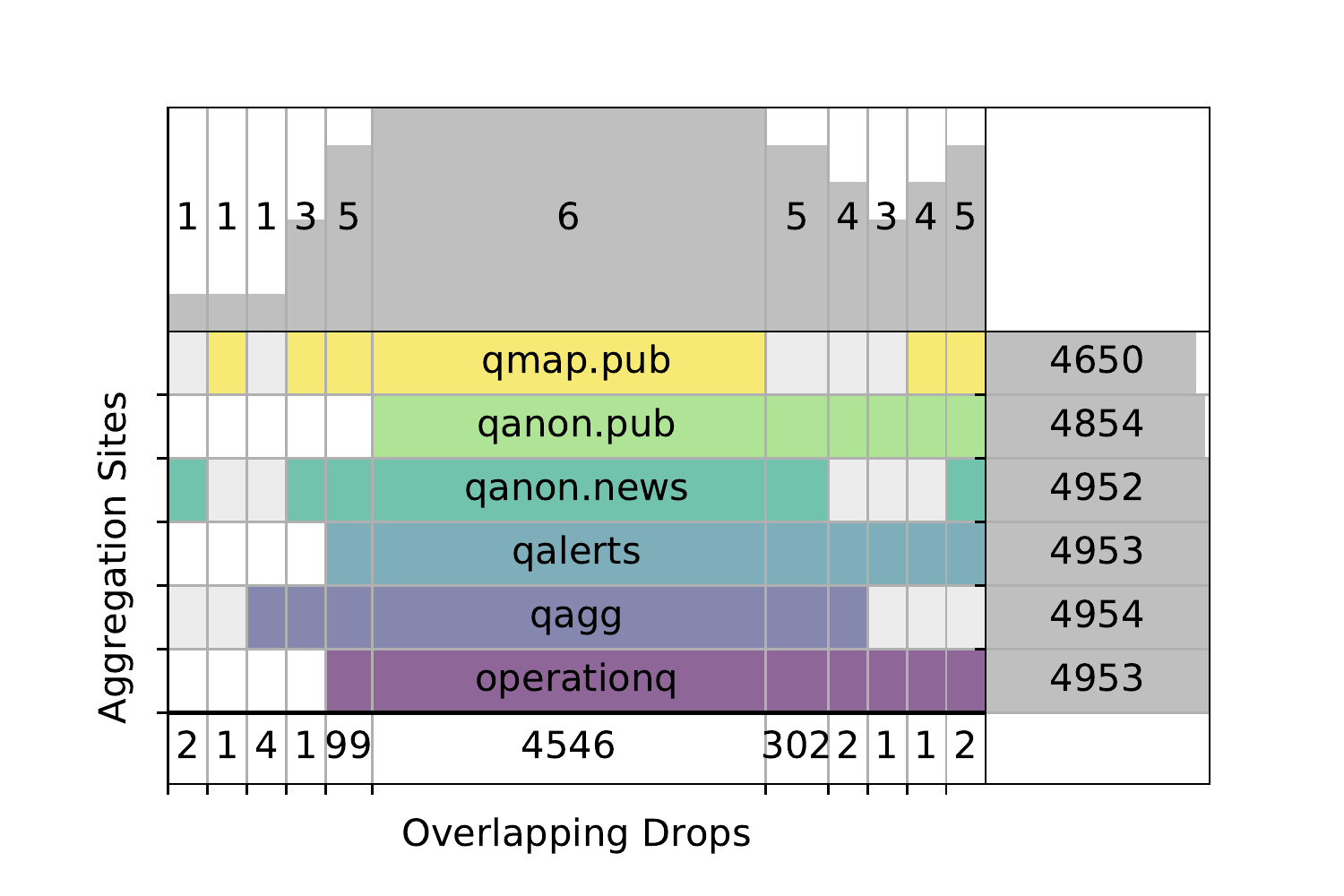}
  \caption{A Venn diagram-esque visualization of discrepancies between Q drops included by different aggregation sites.}
  \label{fig:super_venn_agreement}
\end{figure}

Figure~\ref{fig:super_venn_agreement} depicts the discrepancies in Q drops across our six aggregation sites.
The figure can be read as a Venn diagram, where a cell is shaded for each drop that a given site includes.
Along the bottom of the figure, we show the number of drops unique to the intersection of aggregation sites that are shaded.
We show the number of sites that agree about the given set of drops along the top of the figure.
For example, in the middle of the figure is the set of 4,546 Q drops that all six sites include.

Exploring the discrepancies in this manner reveals a few things.
First, directly to the right of the block of 4,546 drops that all sites include, we find the 302 that all sites except qmap.pub include.
While most of these drops (300) are posted after qmap.pub shut down, two are not.
As we show later (see Section~\ref{sec:drop_spread}), qmap.pub is the most linked aggregation site on social media.
Although these two Q drops are posted in 2018, well before qmap.pub went offline, manual examination does not explain why qmap.pub discards them.
To the left of this block are the 99 Q drops that all sites except qanon.pub, include.
One possibility for these not being included by qanon.pub is that they might have been improperly attributed to a different board or post ID.
Alas, manual examination reveals this not to be the case: these 99 drops do not appear on qanon.pub in any fashion.

As we discuss later (Section~\ref{sec:drop_spread}), qanon.pub is the most popular aggregation site in the early days of QAnon that spread to the mainstream via Reddit. 
So, the absence of these drops has profound implications for the evolution and current state of the conspiracy theory.

\descr{Duplicate content.}
There are several drops with identical text that Q posted over time.
For example, Q posted the text ``\emph{Worth remembering}'' 5 times in July 2019 to prompt adherents to remember the topic discussed in an imageboard thread title or post.
We believe that Q attempts to provide a promise or prophesy using this text as if the topic of discussion will materialize in the real world.
We find 92 instances where Q shares a post shortly before posting the same content using a different tripcode on different boards.
On other occasions (135 times), Q waits several days or months before posting the same content again.
Such content might be an exact duplicate of previous content, and other times it is just a QAnon hashtag (``WWG1WGA'') or the ``Q'' signature alone.
We find 31 duplicate posts in our dataset, which occur 227 times according to the aggregation sites.
We believe these duplicates have little effect on the remaining analysis as they are only $4.5\%$ of our Q drop dataset, and $74\%$ (168) of them are less than 10 characters long (either a hashtag or the ``Q'' signature).

\descr{Missing content.}
We manually search our imageboards dataset for posts that use Q tripcodes to confirm that the Q drops reported by the aggregation sites did take place.
This returns 299 posts that use Q tripcodes, but are not included in any aggregation site; 247 were posted in March 2018 alone.
163 of these missing Q drops are written by tripcode \texttt{C} (see Table~\ref{table:tripcodes} for the full tripcode).
In an effort to understand why these missing posts are not included in aggregation site lists, we read dozens of these apocryphal posts, finding that the discussion that took place in March 2018 was between tripcode \texttt{C} and tripcode \texttt{E} fighting each other over who was the real Q (e.g., {\em ``You failed miserably fake Q! [...] Go to hell Q-Larp!,'' ``Q came in and created a psy-op [...], '' ``Q sez he's real Q but can't post as real Q! [...].''}

Apparently, many tripcodes were cracked in March 2018 and in August 2018 the passwords were released in an online forum~\cite{qcracked}.\footnote{See related discussion on Reddit: \url{https://bit.ly/3yeVl2G}}
When the tripcodes got ``compromised,'' tripcode \texttt{A} was created and others that had access to cracked tripcodes were accusing Q of being a ``Q-LARP'' (Live Action Role Playing) and a ``shill'' (a person that pretends to support a conspiracy so that they can spy on conspirators). 
Since those posts are attacking the Q persona and do not provide fundamental elements of the conspiracy, it makes sense that the aggregation sites do not archive them.

Overall, our analysis shows that the {\em source material} QAnon is derived from is not clearly defined.
In part, it relies on the interpretation of aggregation site operators.
These discrepancies across aggregation sites demonstrate holes in a single source of truth of the conspiracy.
While a deeper exploration is outside the scope of this paper, it does show the relative power that aggregation sites have in curating and archiving the otherwise ephemeral ``official'' teachings of Q.

\textbf{NB:} With this caveat, in the rest of the paper, we treat a set of 4,949 unique Q drops included in at least \emph{five} aggregation sites as the canonical set of Q drops.

\subsection{Is There a Canonical Q?}\label{sec:stylometry}
Considering the murky nature of Q in general, we now explore Q's behavior over time.
We compare the posts made by the different tripcodes that have been deemed canonical by at least 5 aggregation sites.
Table~\ref{table:tripcodes} lists all tripcodes, along with the number of posts they made.
We further label each tripcode for visualization purposes in the following analysis.

\begin{table}[t]
\small
\setlength{\tabcolsep}{4pt}
\centering
\begin{tabular}{ c l r }
\toprule
\textbf{Label} & \textbf{Tripcode} & \textbf{\#Posts} \\
\midrule
{\tt A} & !!mG7VJxZNCI & 1,796  \\
{\tt B} & !!Hs1Jq13jV6 & 1,315  \\
{\tt C} & !UW.yye1fxo & 583  \\
{\tt D} & !CbboFOtcZs & 399  \\
{\tt E} & !xowAT4Z3VQ & 351  \\
{\tt F} & !ITPb.qbhqo & 223  \\
{\tt G} & \emph{no tripcode} & 163  \\
{\tt H} & !4pRcUA0lBE & 94  \\
{\tt I} & !A6yxsPKia. & 16  \\
{\tt J} & !2jsTvXXmXs & 9  \\
\midrule
\textbf{Total} & & 4,949\\
\bottomrule
\end{tabular}
\caption{Tripcodes and the number of posts they made.}%
\label{table:tripcodes}
\end{table}

\begin{figure}[t]
    \centering
    \includegraphics[width=0.49\textwidth]{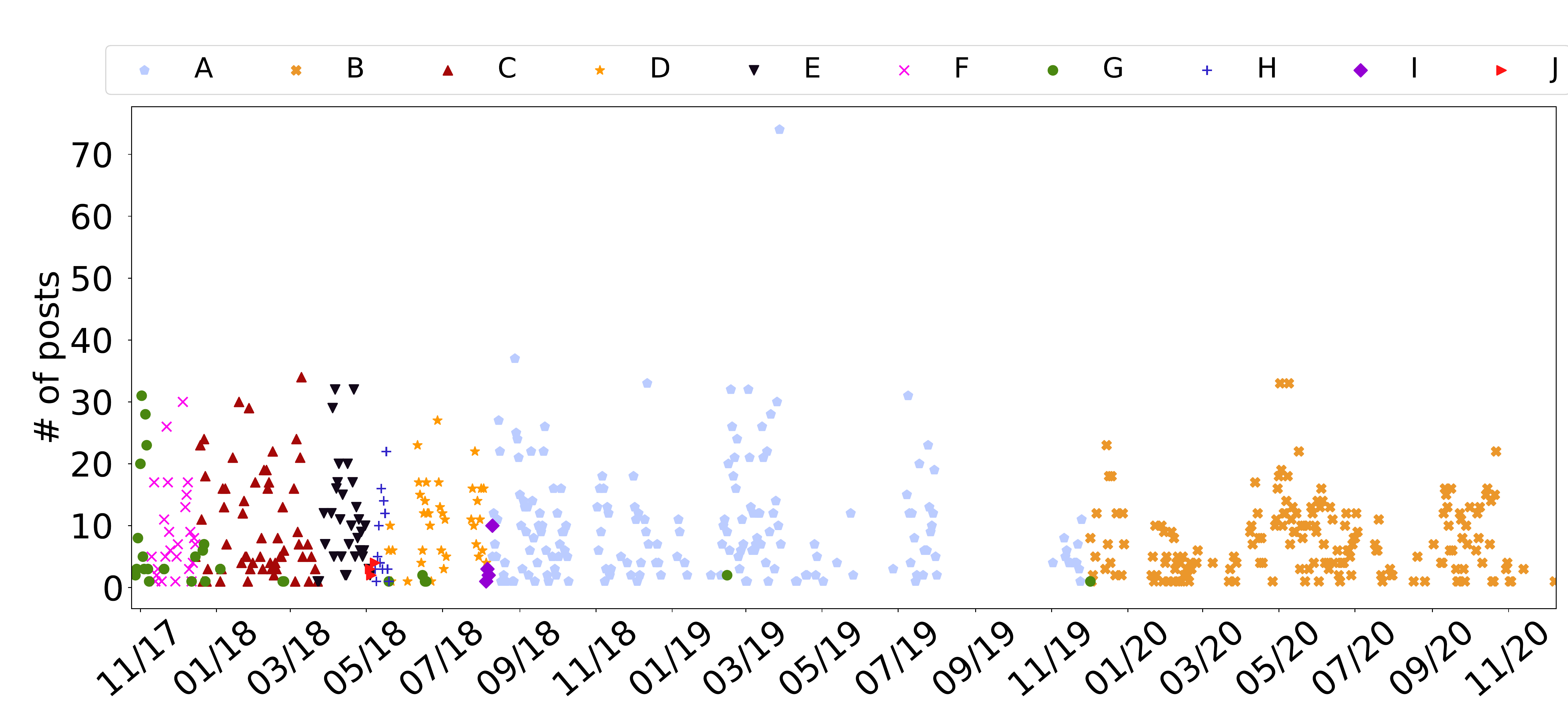}
    \caption{Posts per day per tripcode.}
    \label{fig:tripcode_post_per_day}
\end{figure}

\begin{figure*}[t!]
\center
\subfigure[]{\includegraphics[width=0.35\textwidth]{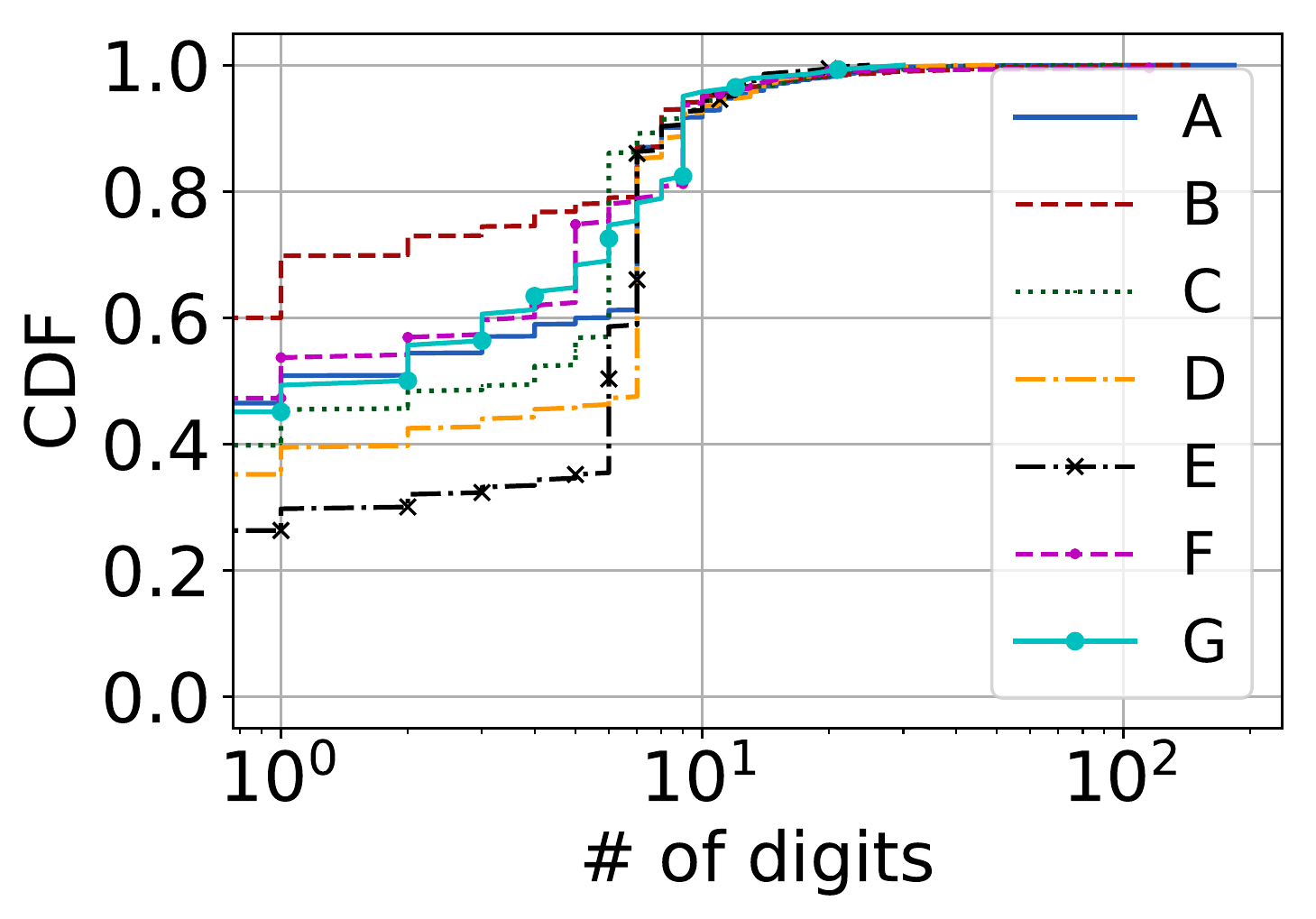}\label{fig:cdf_tripcode_digits}}
~~~
\subfigure[]{\includegraphics[width=0.35\textwidth]{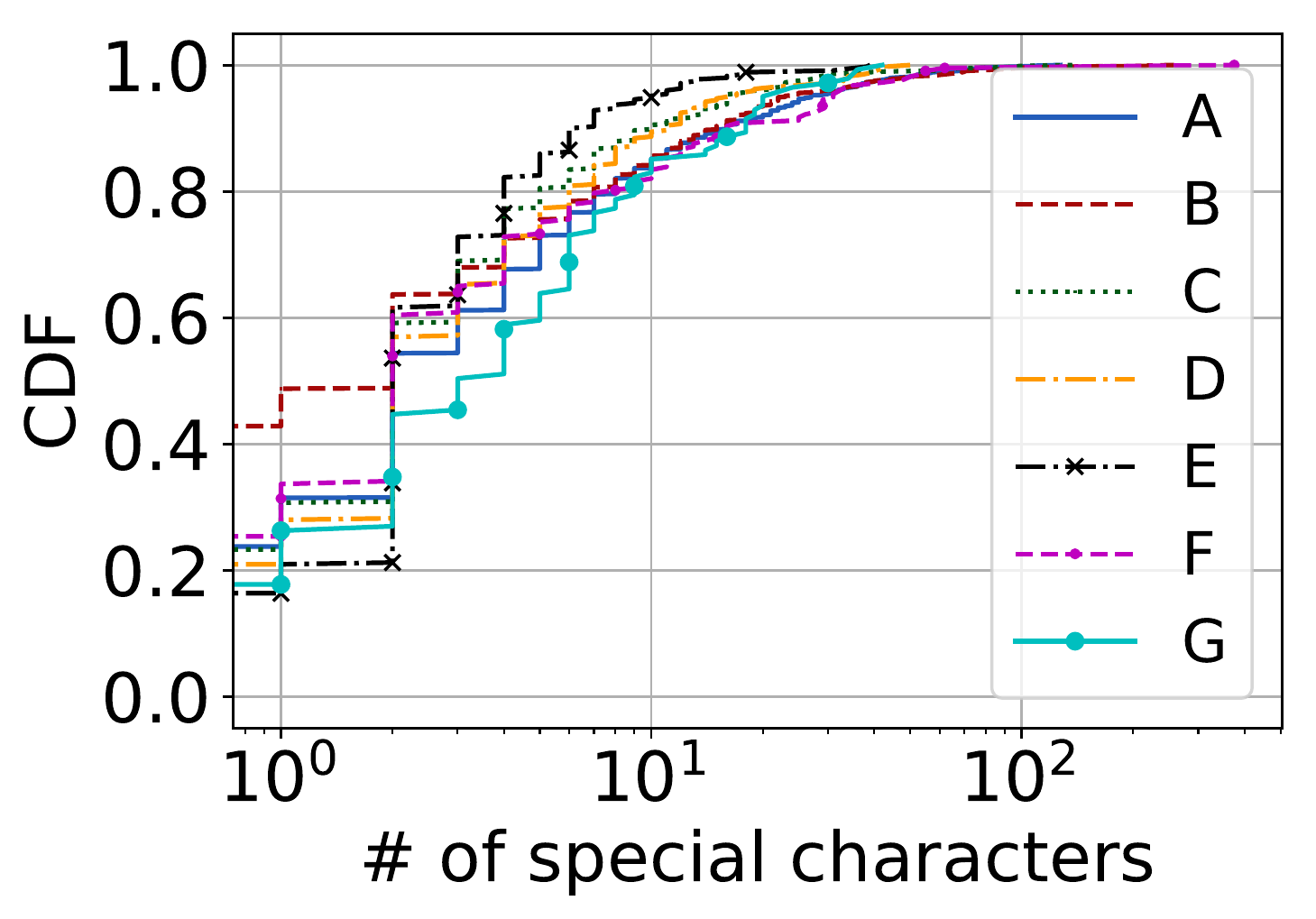}\label{fig:cdf_tripcode_special}}
\caption{CDF of the number of (a) digits and (b) special characters per post for each tripcode.}
\label{fig:tripcode_digit_special}
\end{figure*}

\begin{figure*}[t!]
\center
\subfigure[]{\includegraphics[width=0.35\textwidth]{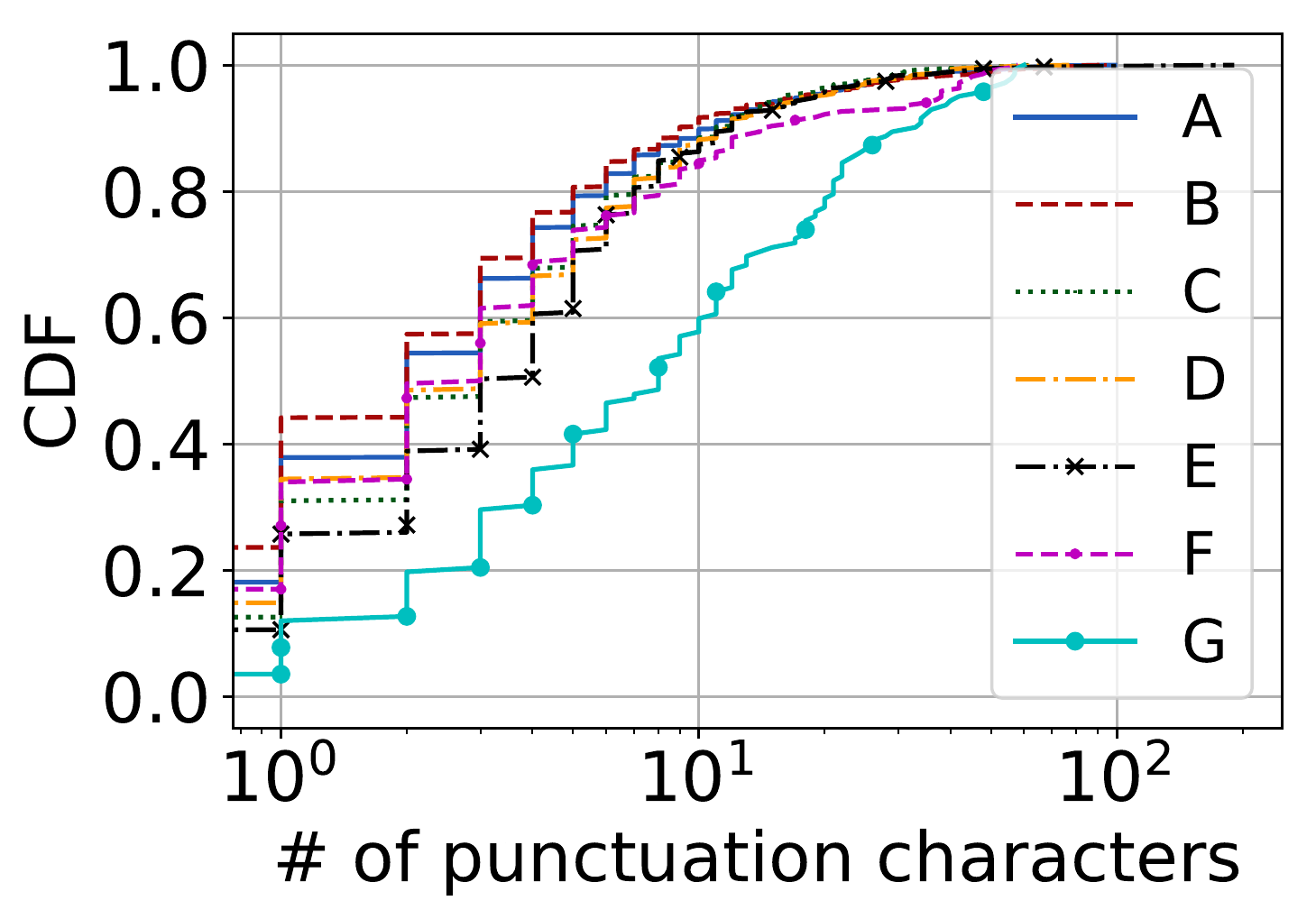}\label{fig:cdf_tripcode_punctuation}}
~~~
\subfigure[]{\includegraphics[width=0.35\textwidth]{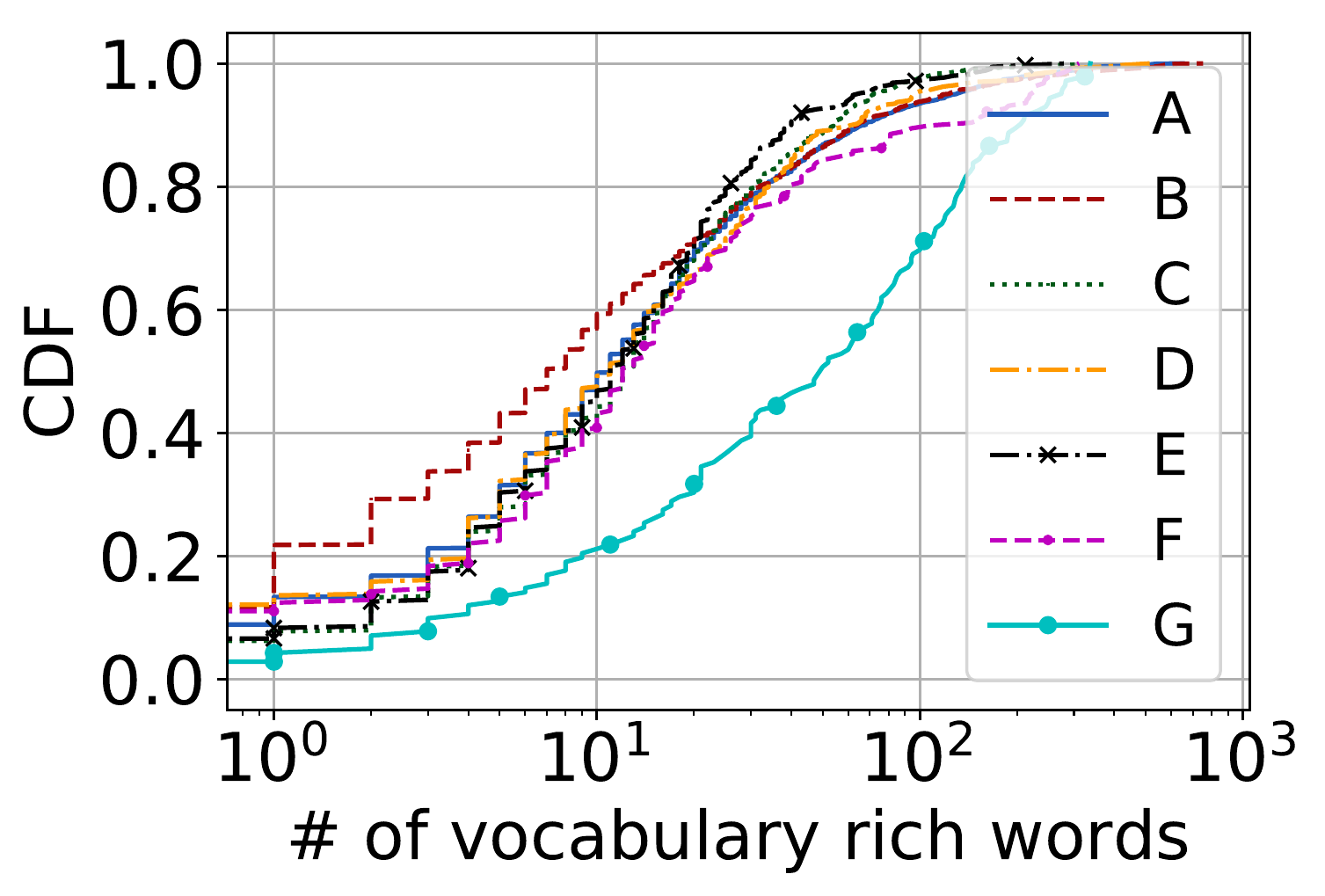}\label{fig:cdf_tripcode_rich}}
\caption{CDF of the number of (a) punctuation characters and (b) vocabulary rich words per post for each tripcode.}
\label{fig:tripcode_punctuation_richness}
\end{figure*}

\descr{Q drops by tripcode.}
In Figure~\ref{fig:tripcode_post_per_day}, we plot the number of posts per day by the different tripcodes attributed to Q.
The majority are used within the first nine months of the conspiracy theory's creation.
Interestingly, this is also the time period where the most overlap between tripcodes occurs -- in the first nine months, the tripcode attributed to Q changed six times, and there are several time periods where drops with no tripcode (i.e, {\tt G} in Figure~\ref{fig:tripcode_post_per_day}) overlap with other drops. 

Notably, after the publication of the passwords of the compromised tripcodes in August 2018~\cite{qcracked}, all tripcodes go silent, and tripcode {\tt A} is created and remains active until 8chan went down in August 2019.
Then, tripcode {\tt A} appears again when 8chan resurfaces as 8kun in November 2019~\cite{8chandown}, before being replaced by tripcode {\tt B}.

The question around authentication/authenticity of Q and the posts they made is central to understanding the conspiracy theory in general.
Simply put, if Q was not the same person across time, it would, for all intents and purposes, be a meaningful blow to the sustained narrative.
Indeed, the overlap in Figure~\ref{fig:tripcode_post_per_day} indicates that additional attempts to disambiguate potentially different authors of Q drops warrants further exploration.

\descr{Q's writing habits.} 
The conspiracy suggests that a government official provides adherents with inside information about the struggle over power between the deep-state and Donald Trump.
One way of exploring the question of whether or not Q is one person is to use stylometry.
In other words, we perform stylometric analysis to identify differences in the way that different tripcodes write.
Existing literature has used hundreds of stylometric features and analysis methods to solve various problems, e.g., author attribution, identification, etc.
At the same time, researchers disagree on which features should be used for each specific problem~\cite{brocardo2013authorship}.

Since our Q drop dataset consists only of posts that aggregation sites consider to be written by Q, we assume that each drop was written by one person and focus on comparing the writing habits of each tripcode. 
Although we do use some of the most widely used stylometric features in this analysis~\cite{lagutina2019survey}, we do not take into account the possibility of organic style change over time as only two of the tripcodes are used for more than three months.
These two tripcodes were created after the previous tripcodes got compromised and their passwords published online.
Also, we do not consider author obfuscation of Q drops from a tripcode as we intend to compare how, if at all, Q drops from different tripcodes differ.
Finally, it is important to note that we make an implicit assumption that individual tripcodes are used by a single person based on Q's multiple claims to be a singular individual.

We find that tripcode \texttt{B} exhibits a significantly different number of words and character. %
We reject the null hypothesis of the 2-sample Kolmogorov-Smirnov test~\cite{massey1951kolmogorov} that the distributions are drawn from the same parent distribution ($p < 0.01$).
Interestingly, we test the distributions of the \#words and \#characters per post, and are unable to reject the null hypothesis when looking at the pairs of tripcode {\tt A} vs {\tt C}, {\tt D}, and {\tt E}; {\tt C} vs {\tt D}, and {\tt E}; and {\tt D} vs {\tt E}.

Figure~\ref{fig:tripcode_digit_special} reports stylometrics related to the use of digits and special characters.
Here, we are able to reject the null hypothesis for the use of digits (Figure~\ref{fig:cdf_tripcode_digits}, $p < 0.01$ for all comparisons).
However, for the use of special characters (Figure~\ref{fig:cdf_tripcode_special}), we are unable to reject the null hypothesis when comparing tripcode {\tt D} to {\tt C} ($p = 0.72$) as well as {\tt D} to {\tt A} ($p=0.33$).
As before, we reject the null hypothesis for tripcode {\tt B} vs the other four top tripcodes.
Finally, in Figure~\ref{fig:tripcode_punctuation_richness}, we plot the CDF of punctuation and vocabulary richness of Q drops per tripcode.
Here, we are unable to reject the null hypothesis for tripcode {\tt C} vs {\tt D} for punctuation ($p = 0.92$), while, for vocabulary richness, when comparing tripcodes {\tt A} vs {\tt D} ($p = 0.83$), {\tt C} vs {\tt D} ($p=0.37$), as well as {\tt E} vs  {\tt D} ($p=0.60$).
Again, we reject the null hypothesis for {\tt B} vs the other four tripcodes. 

\subsection{Take Aways}
Our analysis indicates that the six aggregation sites we focus on do not exhibit high agreement scores for what is considered to be a Q drop.
Therefore, the content echoed by individuals sharing aggregation sites on mainstream platforms could be conflicting, resulting in a different view of the conspiracy theory depending on the information source.
In other words, this high level of disagreement indicates that conspiracy theories can ``thrive'' despite their narratives may at times contradict their core beliefs.
In addition, stylometry shows that there are questions on whether the canonical posts from Q have different authors over time.
In fact, at the very least, the characteristics of tripcode \texttt{B} are quite different (and in a statistically significant way) from the other top four tripcodes. 
Also, note that tripcode {\tt B} only started posting when QAnon moved to 8kun (i.e., November 2019).

Overall, our analysis provides a strong indication that the Q persona was adopted by more than one person, not necessarily in a coordinated manner.
Tripcodes overlap several times, and writing habits change significantly between different ones.
Many individuals came forward to claim they are the person behind Q, but there was no way to prove their claims~\cite{insiderQanalysis}.
That being said, it is likely that these individuals were fascinated by this conspiracy and were interested in sharing their own beliefs within these forums, using a signature (Q) that everyone would notice.

\section{Q Conspiracy Analysis}\label{sec:qanalysis}
In this section, we analyze the content of the Q drops, exploring the topics discussed, and investigating the nature of the content, e.g., its toxicity, coherence, etc.

\subsection{What does Q discuss?}\label{sec:topics}
Considering the prominent nature of QAnon in real-world events, understanding what Q actually talks about is particularly relevant, and so is discovering the topics a cult of adherents has formed around.

\descr{Word Embeddings.}
To assess how different words are interconnected within the Q drops, we use word2vec, a two-layer neural network that generates word representations as embedded vectors~\cite{mikolov2013efficient}.
This model takes a corpus of text and maps each word to a multi-dimensional vector in a linear space.
This means that words used in similar contexts tend to have similar (``close'') vectors.

We use word2vec to detect the main topics of discussions in Q drops, rather than alternatives like Latent Dirichlet Allocation (LDA) or Biterm Topic Model, as the latter return incoherent topics with significantly low coherence scores (we omit details to ease presentation).
Previous studies show that this approach works well on corpus similar to ours, e.g., tweets~\cite{vargas2019characterization}, and Voat posts~\cite{papasavva2021qoincidence}.

To clean our corpus we remove all formatting characters, e.g., \textbackslash n, \textbackslash t, and URLs from each Q drop.
We also remove Q's ``signature'' at the end of the drops, %
as well as all numbers, with the exception of numbers included in words (e.g., ``wwg1wga'').
Finally, we tokenize every post and remove stop words.
In the end, we build a corpus of 3.7K drops consisting of 77.8K tokens.
We train our model using a context window (which defines the maximum distance between
the current word and predicted words when generating the embedding) of five, as previous work suggests it is commonly used to capture broad topical content~\cite{levy2014dependency}.
Finally, we limit our vocabulary to words that appear at least ten times because of the small size of our dataset, which yields a vocabulary of 1,673 words.

\descr{Discovering important phrases.}
To identify the most important words in our vector space, we look at the top ten words closest to the \emph{centroid} in the embedding's vector space.
We do so as words closest to the centroid vector tend to be related to the main topics of the corpus; see Rossiello et al.~\cite{rossiello2017centroid}.

The ten words closest to the centroid, along with the computed similarity score of the words and centroid embeddings, are: throw (0.737), laying (0.705), jim (0.703), despotism (0.649), priestap (0.648), importance (0.640), judiciary (0.634), heavenly (0.626), independent (0.625), evinces (0.615).
A manual examination of the Q drops indicates that these are indeed common topics of discussion.
For instance, Q promotes the over-{\em throw} of the government, which is allegedly run by despots, and the institution of a new one. 
Also, Q speaks often about law enforcement and criminal justice figures like E. W. Priestap (an attorney) and Jim Rybicki (formerly at the US Dept. of Justice).

\descr{Word visualization and topics.}
We visualize broader topics using the methodology of Zannettou et al.~\cite{zannettou2020quantitative}, transforming the embeddings into a graph, where nodes are words and edges are weighted by their cosine similarity to other words.
This requires removing edges from the graph that are below a certain threshold of similarity (i.e., we prune low similarity edges).
We use a threshold of 0.6 based on the distribution of pair-wise vector similarities from our embeddings (we omit the figure due to space constraints) and for visualization purposes. 
Finally, we perform community detection using the Louvain method~\cite{blondel2008fast} on the resulting graph to provide insights into the high-level topics that individual words form.

\begin{figure*}[t]
    \centering
    \includegraphics[width=0.8\textwidth]{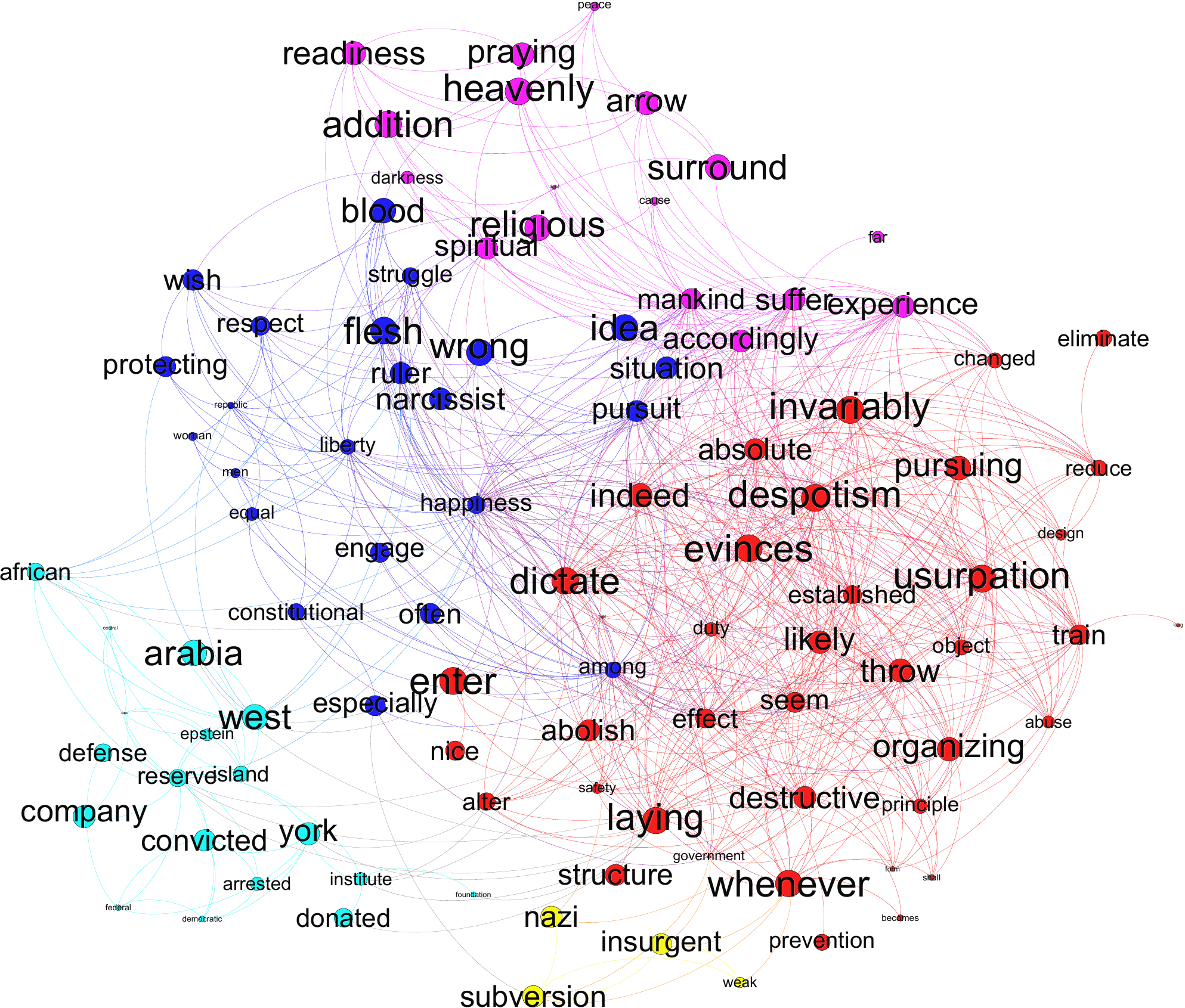}
    \caption{Graph representation of the two-hop ego network, starting from the keyword ``throw.''}
    \label{fig:graph_throw}
\end{figure*}

\descr{Insurgent Communities.} 
Figure~\ref{fig:graph_throw} shows the two-hop ego network around the most similar word to the centroid, i.e., ``throw''.
Similar to Zannettou et al.~\cite{zannettou2020quantitative}, nodes in the figure are colored by the community they form and the graph is laid out using ForceAtlas2~\cite{jacomy2014forceatlas2}, which positions nodes in the two-dimensional space based on the weight of the edges between them (i.e., more similar words are closer together in the figure).
The community that ``throw'' belongs to (in red on the right side of the figure) is related to governments, the alleged despotism they are engaging in, and the supposed duty that patriotic citizens have in addressing these issues.
For instance, ``government,'' ``usurpation,'' ``duty,'' and ``abolish'' all appear in this community.
The small yellow community close to ``government'' specifically discusses ``subversion'' and ``insurgency.''

\descr{Religion.} 
The magenta community at the top of the figure is related to ``religion'' and ``spirituality.''
Interestingly, beneath the ``religion'' community, we find a blue community which discusses ``narcissist'' ``rulers,'' along with ``struggle,'' ``blood,'' and ``flesh.''
These two communities are interconnected as the movement believes that the rulers of the so-called deep-state drink the blood of children in satanic religious rituals.
Finally, the turquoise community at the bottom left of the figure seems to be discussing Epstein and his island, as well as companies and institutions.

\begin{figure*}[t!]
\center
\subfigure[toxicity]{\includegraphics[width=0.32\textwidth]{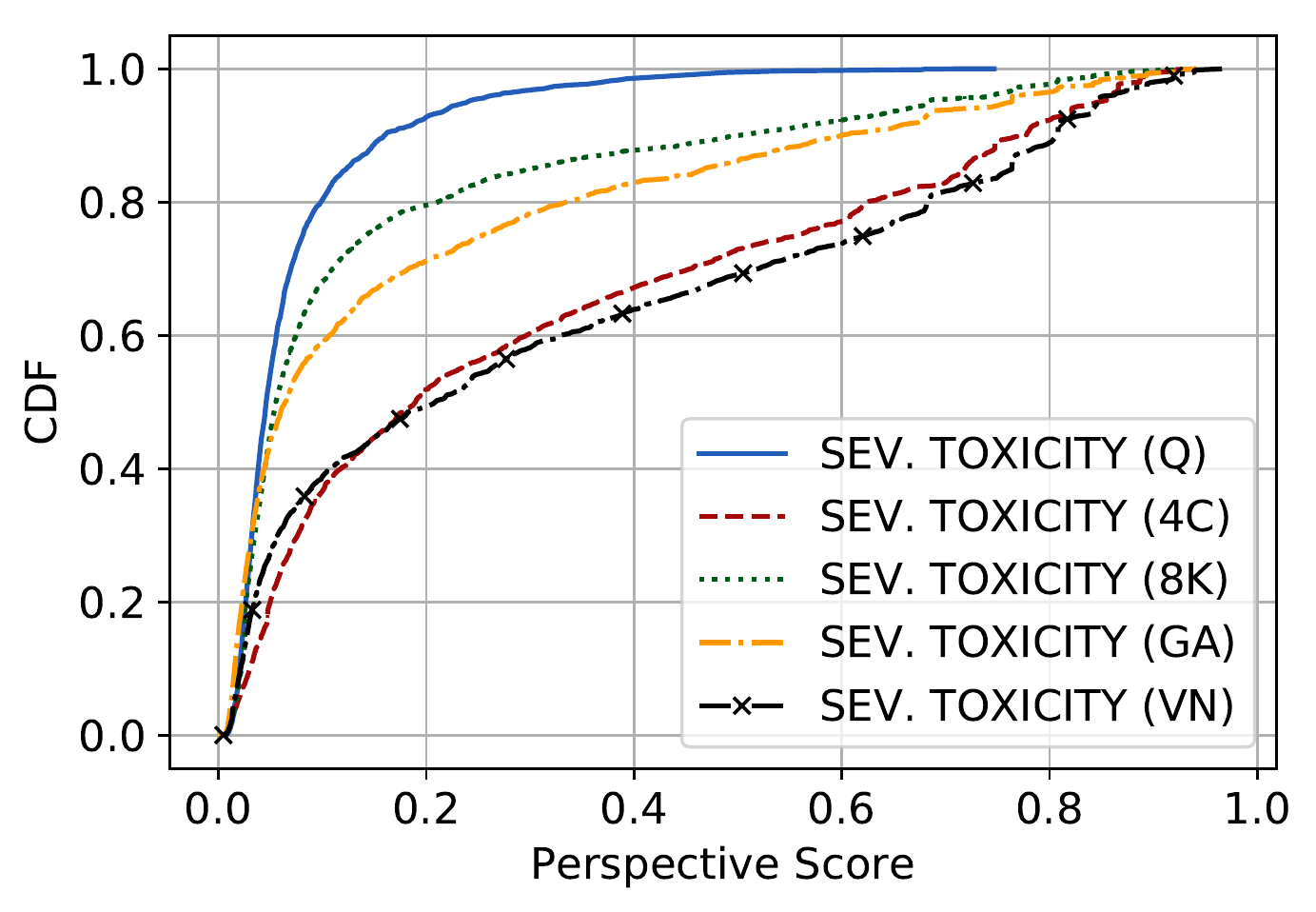}\label{fig:cdf_severe_toxicity}}
\subfigure[threatening]{\includegraphics[width=0.32\textwidth]{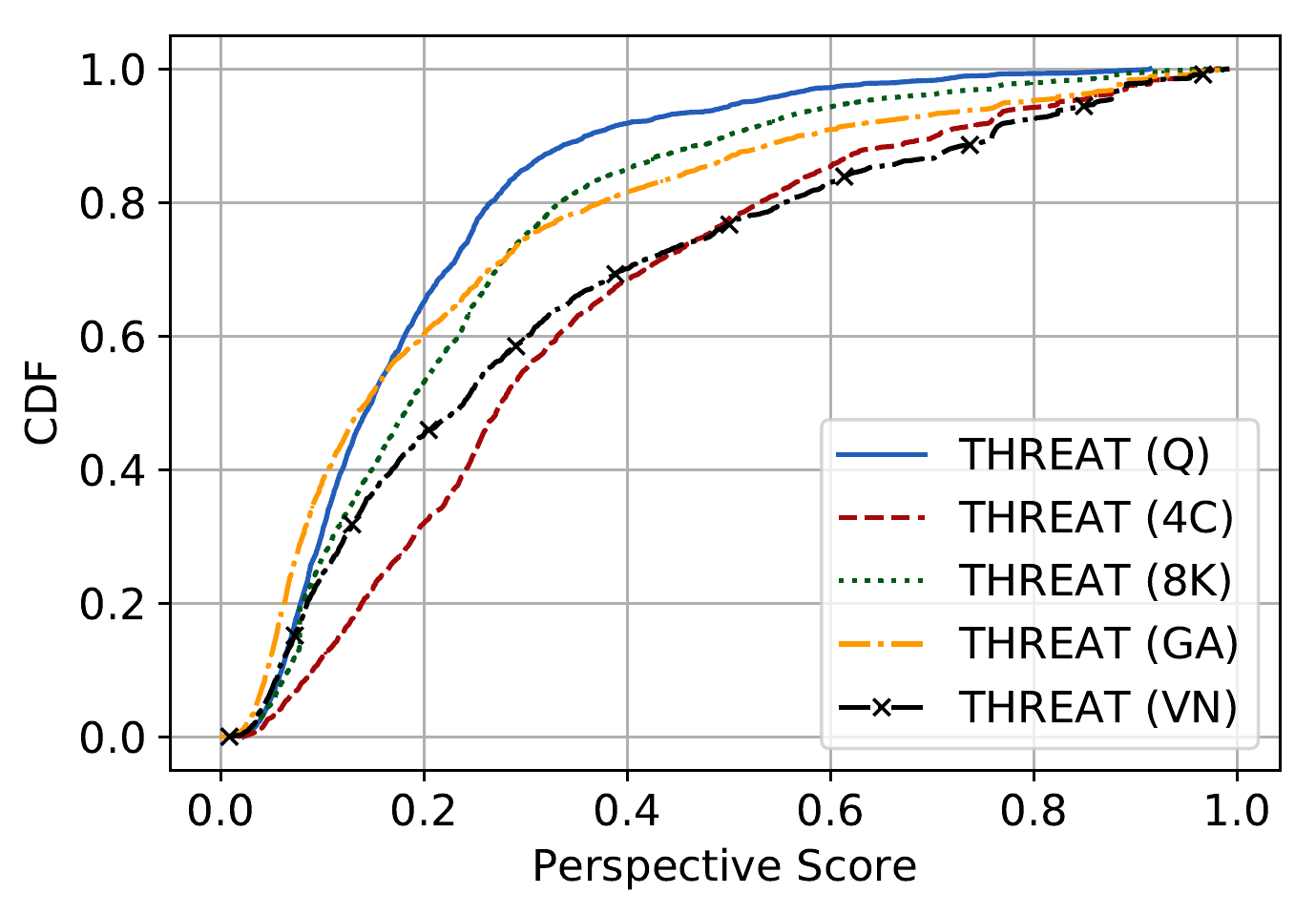}\label{fig:cdf_threat}}
\subfigure[incoherent]{\includegraphics[width=0.32\textwidth]{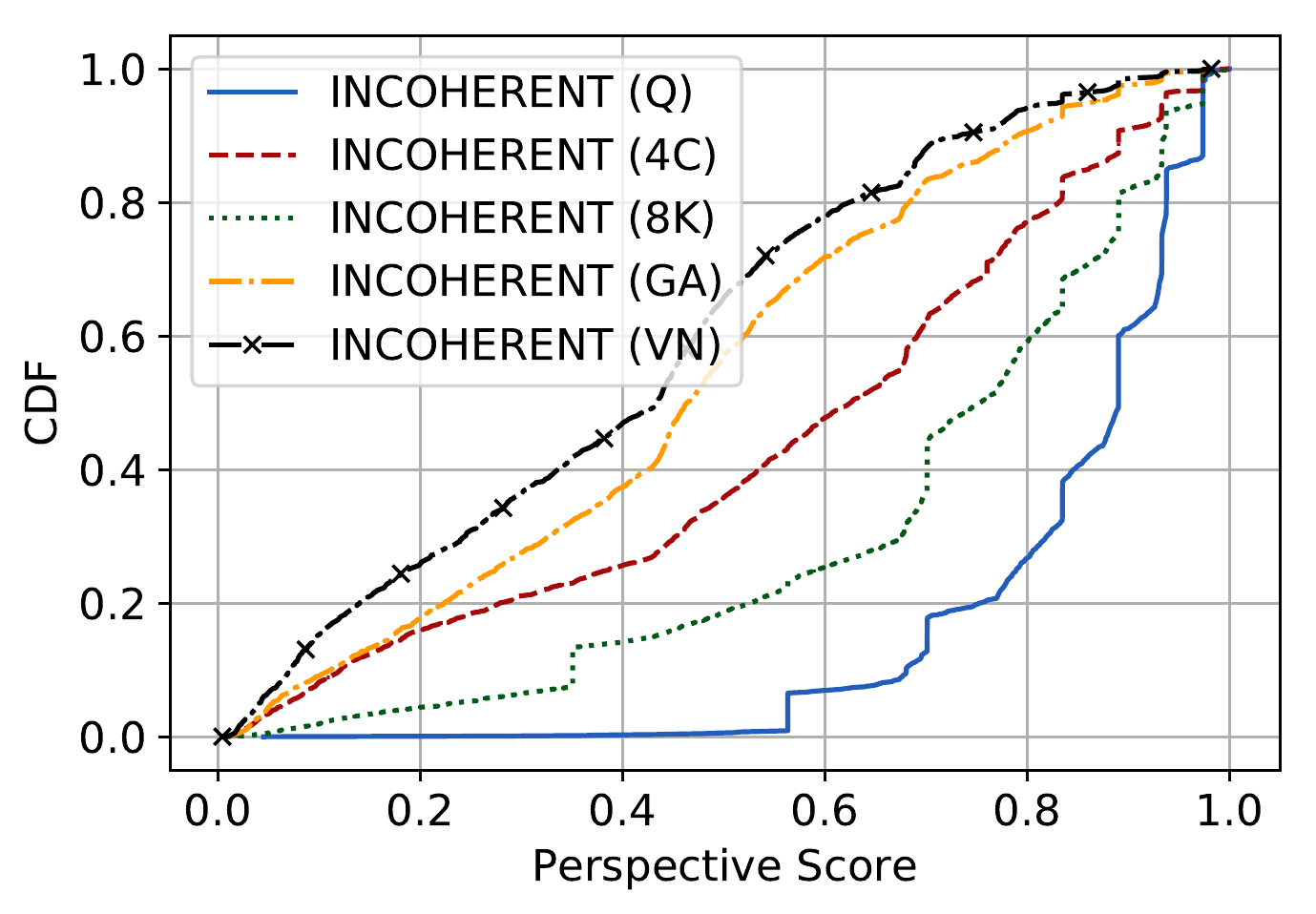}\label{fig:cdf_incoherent}}
\caption{
CDF of the Perspective API scores. ``Q'' stands for Q drops, ``4C'' for 4chan \dspol, ``8K'' for 8kun \qresearch, and ``GA'' and ``VN'' for Voat \greatawakening and \news, respectively.}
\label{fig:cdf_qdrop_perspective}
\end{figure*}

\subsection{Perspective Analysis}\label{sec:toxicity}
More specifically, we use three models made available via Google's Perspective API~\cite{jigsaw2018perspective}: 
1)~ {\em severe toxicity,}
2)~{\em threatening,}
and 3)~{\em incoherent} language.\footnote{For more details, see \url{https://developers.perspectiveapi.com/s/about-the-api-attributes-and-languages}}

These return a score between 0 and 1 and work on text only.
We choose Google's Perspective API to detect hate speech in Q drops as it provides different models to detect the perceived impact a comment may have in a conversation. 
More specifically, we are interested in understanding how \emph{threatening} the Q drops are, since they promote the overthrow of the government, as discussed in Section~\ref{sec:topics}, and were in fact weaponized in the context of the US Capitol storming on Jan 6, 2021~\cite{capitolriot2}.
We also rely on the \emph{severe toxicity} model of the API to show to what extent the Q drops are toxic, which might indicate that their content promotes hate towards specific individuals or groups of people. 
Finally, motivated by the low coherence of the LDA topic detection model (discussed in Section~\ref{sec:topics}), we use the \emph{incoherence} model provided by the API to check whether this model comprehends the content of these posts.

We acknowledge that the use of Perspective API is not without limitations. 
Specifically, previous work has shown that users can evade toxicity detection via simple deception techniques~\cite{hosseini2017deceiving}, while Sap et al.~\cite{sap2019risk} note that the API is biased against posts written in African-American English.
However, we do not take the scores at face value but instead, use them to compare Q drops to text written by other relevant communities.
We are also faced with a lack of alternatives; in fact, Zannettou et al.~\cite{zannettou2020measuring} find that the ``severe toxicity'' model outperforms other tools like HateSonar, while Rajadesingan et al.~\cite{rajadesingan2020quick} show that the ``toxicity'' model of the API yields comparable performance as manually annotated Reddit data.

About 25.1\% (1.2K) of our Q drops either include no text or only have links and/or images. %
We compare the scores of the remaining 3.7K Q drops to those of an equal number of randomly selected posts from 4chan's \dspol, 8kun's \qresearch, and Voat's \greatawakening and \news.

In Figure~\ref{fig:cdf_qdrop_perspective}, we plot the CDFs of the scores for each  model.
The Q drops do not seem to be severely toxic (Figure~\ref{fig:cdf_severe_toxicity}), with a median value ($0.04$), similar to \greatawakening ($0.06$) and \qresearch ($0.05$), but lower than \news and \dspol ($0.2$ and $0.19$, respectively).

Q drops seem to score similar threat median scores (Figure~\ref{fig:cdf_threat}) as Voat's \greatawakening ($0.14$), %
but much lower than \dspol and \news ($0.27$ and $0.24$, respectively). 
Considering what discussed in Section~\ref{sec:topics} related to government overthrow, one would expect the scores of the ``threat'' model to be higher.
Thus, we manually inspect our dataset and find that Q does not tend to directly threaten to harm individuals or groups, which is what the model detects.

Last, Q drops seem to be incoherent, much more so than \dspol, \qresearch, and \greatawakening posts (Figure~\ref{fig:cdf_incoherent}).
Specifically, $99\%$ of the Q drops receive incoherent scores greater than $0.5$; as discussed in Section~\ref{sec:54}, incoherence is noticeable upon manual examination.
We also test for significant statistical differences across all five distributions for all three models, and reject the null hypothesis for all distributions ($p<0.001$).

\subsection{Conspiracy Spread}\label{sec:drop_spread}

The prevailing thought on how QAnon gained widespread adherents was that several actors were responsible for spreading it from fringe imageboards to the mainstream Web, creating accounts, and curating communities endorsing and promoting the conspiracy theory on Reddit, YouTube, and Twitter~\cite{smith2020}.
In particular, anecdotal evidence suggests that Reddit played a vital role in QAnon's transition to mainstream adoption~\cite{zadronzny2018}, although it was also the first to exercise related content moderation policies~\cite{redditTOvoat2}.

This prompts us to examine the activity in QAnon-focused subreddits to understand how QAnon content spread on Reddit.
We also analyze our datasets across several axes to shed light on how QAnon was disseminated overall.
Measuring and understanding the way that QAnon spread to the mainstream allows researchers and Internet safety advocates to learn invaluable lessons for the future.
Although our research explores what happens on Reddit after the fact, we still provide insight into characterizations of the conspiracy theory on a mainstream social network, as well as the impact of on-platform enforcement actions.

\begin{figure}[t!]
  \centering
  \includegraphics[width=\columnwidth]{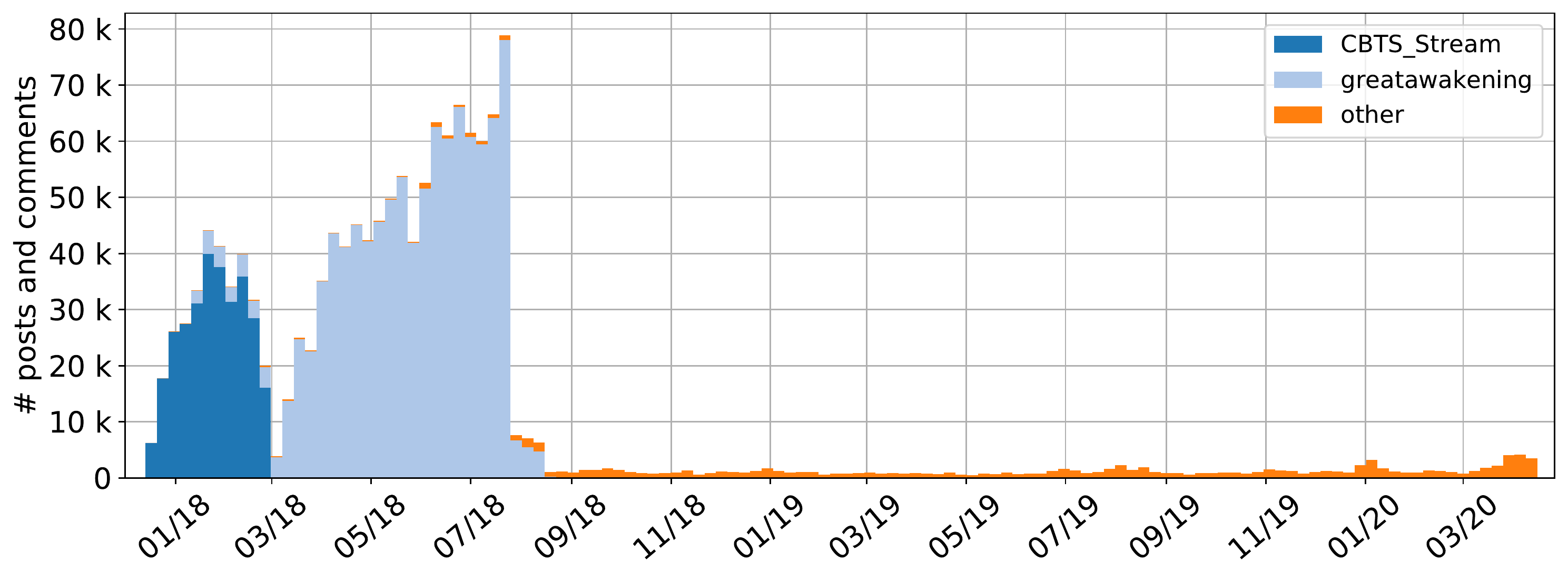}
  \caption{Unique number of posts and comments per week across QAnon related subreddits.}
  \label{fig:reddit_q_sub_activity}
  \end{figure}

\descr{Reddit Activity.} 
In Figure~\ref{fig:reddit_q_sub_activity}, we plot the total content (submissions and comments) for r/CBTS\_Stream and r/greatawakening, created in November 2017 and January 2018, respectively, as well as for the remaining 11 QAnon-related subreddits in our dataset combined (``other'' in the figure).
Note that  r/CBTS\_Stream was the first major subreddit focused on QAnon, which saw an explosive growth in content in late 2017.
The activity starts to decline in February 2018, and eventually ceases on March 14th 2018, when it was banned by Reddit for inciting violence~\cite{wyrich2018}.

Although there was some content posted in r/greatawakening, this subreddit was essentially unused until r/CBTS\_Stream was banned, at which point, over the course of 7 weeks, it exceeded its volume.
At its peak, right before it was banned in early September 2018, r/greatawakening had reached twice the volume of the r/CBTS\_Stream peak. 

Since the banning of these two subreddits, QAnon-related activity on Reddit is reduced significantly; the combined activity of the remaining 11 subreddits is minuscule in comparison.
In fact, the only regularly active subreddit is currently r/Qult\_Headquarters, a community focused on {\em debunking} QAnon.
This follows the general trend suggesting that hard moderation of troubling communities does manage to clean up the platform when appropriate action is taken, however, banned users will probably migrate to other communities~\cite{moderationOnOSN}.
Because of its super conspiracy nature and close relationship to extremist ideology, Q drops may still be disseminated.

\begin{figure}[t!]
  \centering
\includegraphics[width=\columnwidth]{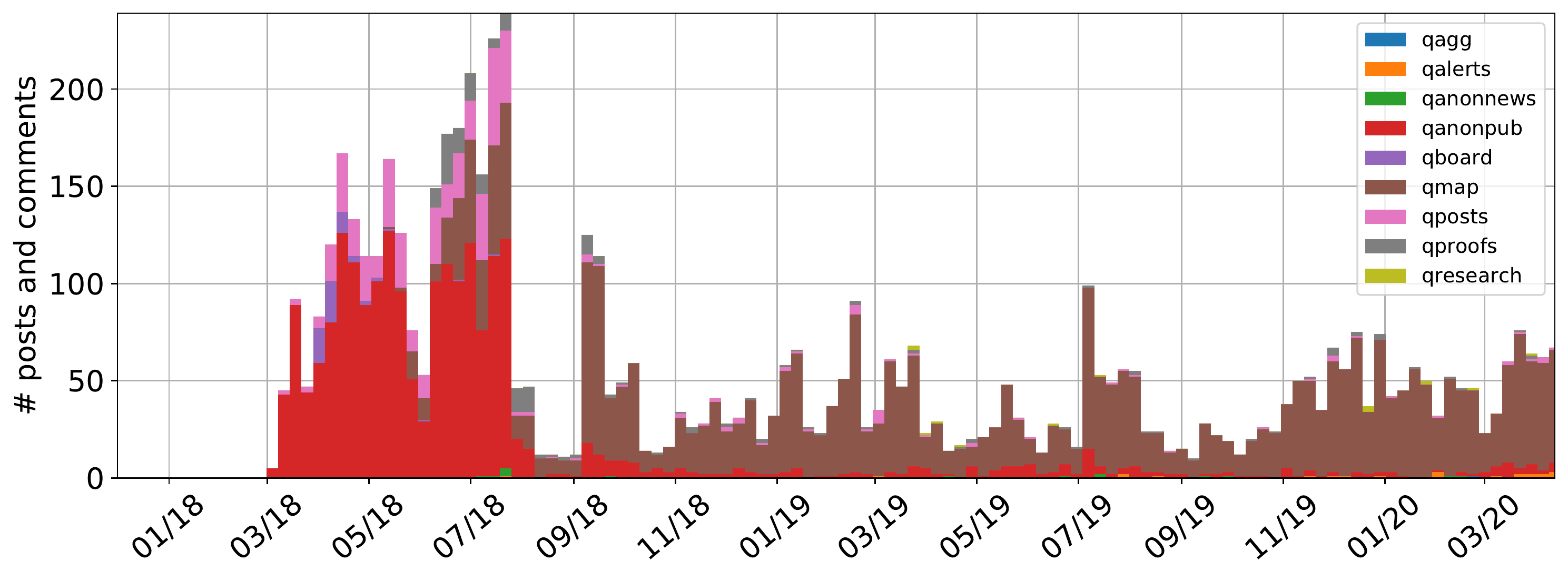}  \caption{Number of posts and comments mentioning the different aggregation sites across all of Reddit.}
\label{fig:reddit_unique_posters_mentioning_agg_sites}
  \end{figure}

\begin{figure}[t!]
\includegraphics[width=\columnwidth]{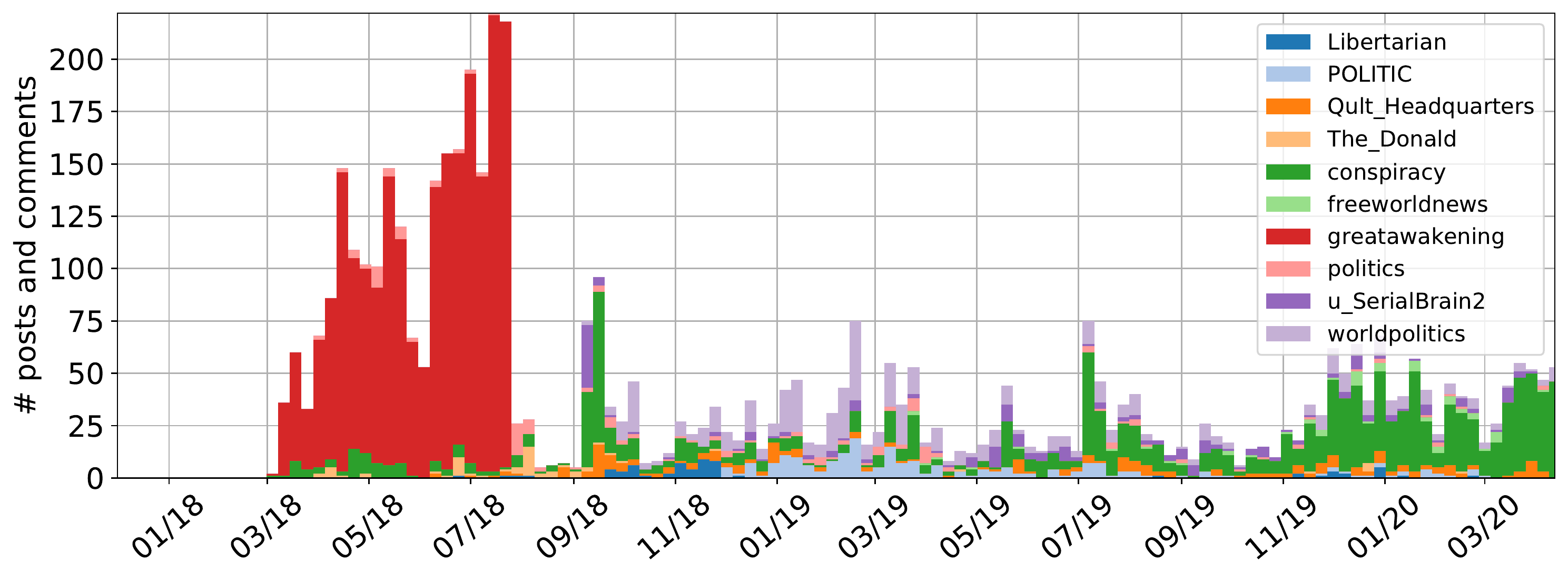}
  \caption{Aggregate site mentions on Reddit broken up by the top 10 subreddits.}
  \label{fig:reddit_agg_sites_by_subreddit}
  \end{figure}

\descr{Links.} 
In addition to specific QAnon-related subreddits, we also explore links to aggregation sites as an indication of whether the conspiracy theory spread on Reddit.
We use these links to measure shared QAnon content on Reddit, especially after the subreddit bans.

We plot the occurrences of links to Q drop sites on Reddit in Figure~\ref{fig:reddit_unique_posters_mentioning_agg_sites}.
When looking at the individual aggregation sites that are linked to, we notice that r/CBTS\_Stream was not responsible for disseminating any Q drop links, as it was banned before the first appearance of a link to an aggregation site.
Domain registration information shows that qanon.news was registered in December 2017, and was the only aggregation site existing prior to r/CBTS\_Stream ban.
Considering that links to aggregation sites began to appear as soon as r/greatawakening received the traffic of r/CBTS\_Stream, we speculate that these sites were created as a response to r/CBTS\_Stream's ban. 
It is common that content banned from mainstream online social networks will result in another dedicated site or social network to host the policy-violating material.
For example, when Reddit banned various hateful communities in June 2015, these communities reemerged on Voat~\cite{redditTOvoat2}.
This implies that although platform enforcement is often successful at removing the content from their own site, it does not thwart actors from spreading it elsewhere.

Although there is diversity in the use of different aggregation sites, the majority of links are to qanon.pub and qmap.pub.
The former was primarily used during the period that r/greatawakening was active, but since then, the latter became the favored aggregation site until it was eventually shut down.
We believe that qmap.pub's rise in popularity was, in large part, due to a dedicated mobile app and bits of content (e.g., QAnon-related definitions and news) that other aggregation sites do not have.

We also focus on the top ten subreddits that links were posted to (Figure~\ref{fig:reddit_agg_sites_by_subreddit}) and find similar levels of diversity: QAnon-related subreddits were not the only ones to share aggregation links. 
While dominated by r/greatawakening, r/conspiracy has consistently posted links to Q drops, and this trend is increasing towards the end of our dataset.

Next, even though r/Qult\_Headquarters is the most active QAnon oriented subreddit remaining, it has relatively few Q drops linked.
Our understanding is that links to Q drops are primarily used by r/Qult\_Headquarters users to point out contradictions or help untangle interpretations by adherents.
Finally, we note the appearance of the left-leaning r/politics in stark contrast to the remaining subreddits, which are largely right-wing subreddits known for extremism and racist ideology~\cite{libertarian,TD}. 

The majority of aggregation site links posted on Reddit are on subreddits supporting the conspiracy, later banned from the platform.
There are some instances where aggregation sites were posted on subreddits like ``PoliticalHumor,'' a subreddit where the main topic was not QAnon: we manually investigated dozens of these posts and observed that majority of them are instances where QAnon conspirators attempt to convince readers of the conspiracy theory.
This suggests that the conspiracy community concentrates on a few supporting subreddits, making a fairly minimal effort to spread itself outwards through aggregation links.
Focusing on removing the communities responsible for posting these links results in significantly slowing aggregation link content sharing on Reddit overall.

\begin{figure}[t!]
\centering
\includegraphics[width=0.77\columnwidth]{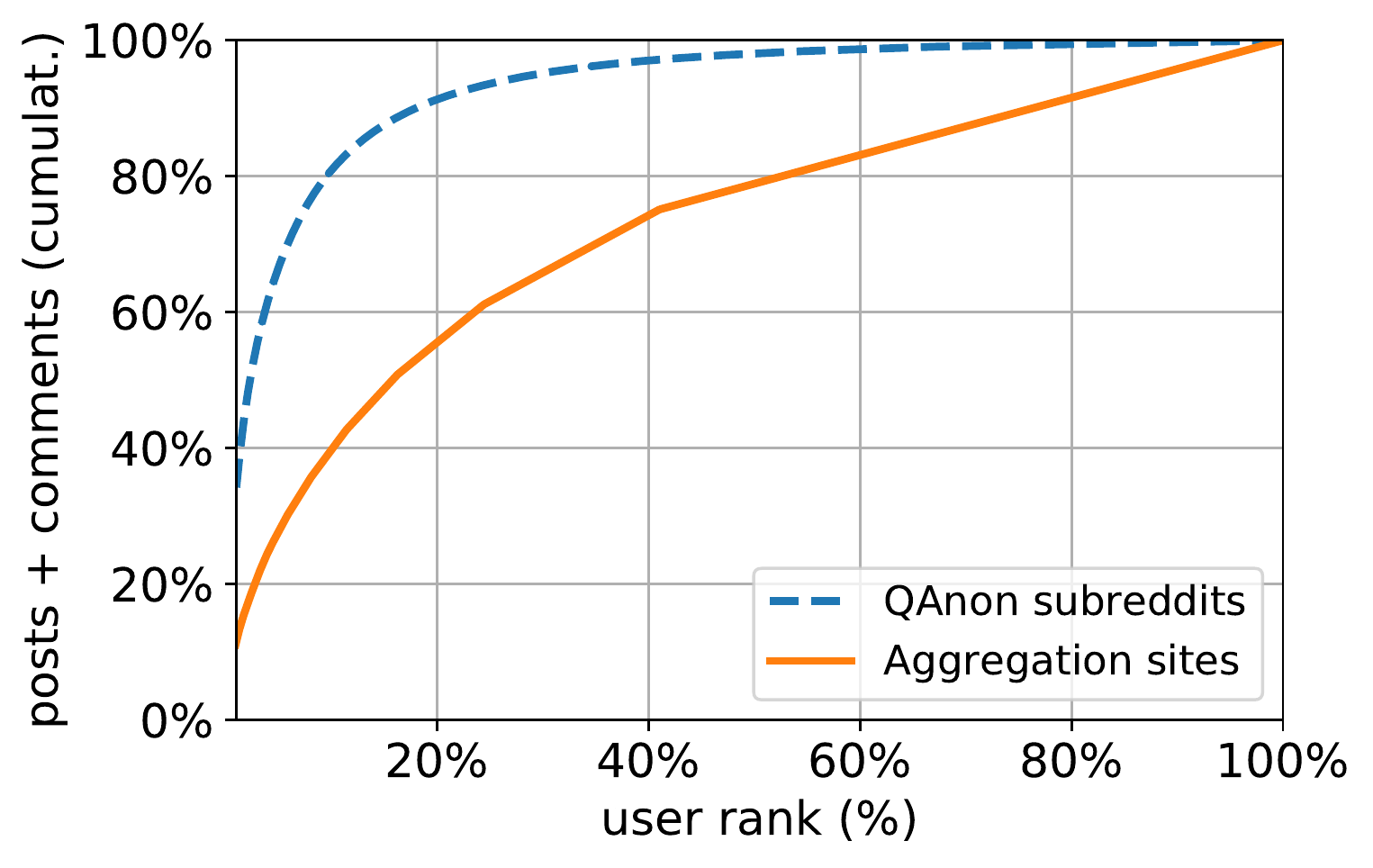}
  \caption{Rank plot showing the percent of Reddit comments made by percent of users.}
  \label{fig:rank_reddit_commenters}
  \end{figure}

\descr{Users.}
Finally, Figure~\ref{fig:rank_reddit_commenters} provides a rank plot with the percentage of the user base making a corresponding percentage of the comments.
For example, we find that $20\%$ of users made over $90\%$ of the comments on QAnon subreddits, suggesting that a few prominent individuals control the conversation.
This is similar to Voat, as submissions in \greatawakening are made by only 346 users out of the 20K subscribers~\cite{papasavva2021qoincidence}.
Over $90\%$ of users made ten or fewer comments mentioning aggregation sites, which, considering the volume of comments in Figure~\ref{fig:reddit_agg_sites_by_subreddit}, is a clear indication that aggregation site postings were done by a select few users. 
In fact, the most prolific user shared links over one thousand times; more than six times the following highest account.

From a platform policy perspective, taking enforcement actions on the small core policy-breaking users would result in a large amount of platform clean up. 
This may contrast with the common platform message that QAnon specifically was ``too large to quickly handle.'' 
However, this shows that removing a small core set of users sharing links to aggregation sites would dramatically reduce the conspiracy theory's spread on the platform.

\begin{figure}[t]
  \centering
  \includegraphics[width=\columnwidth]{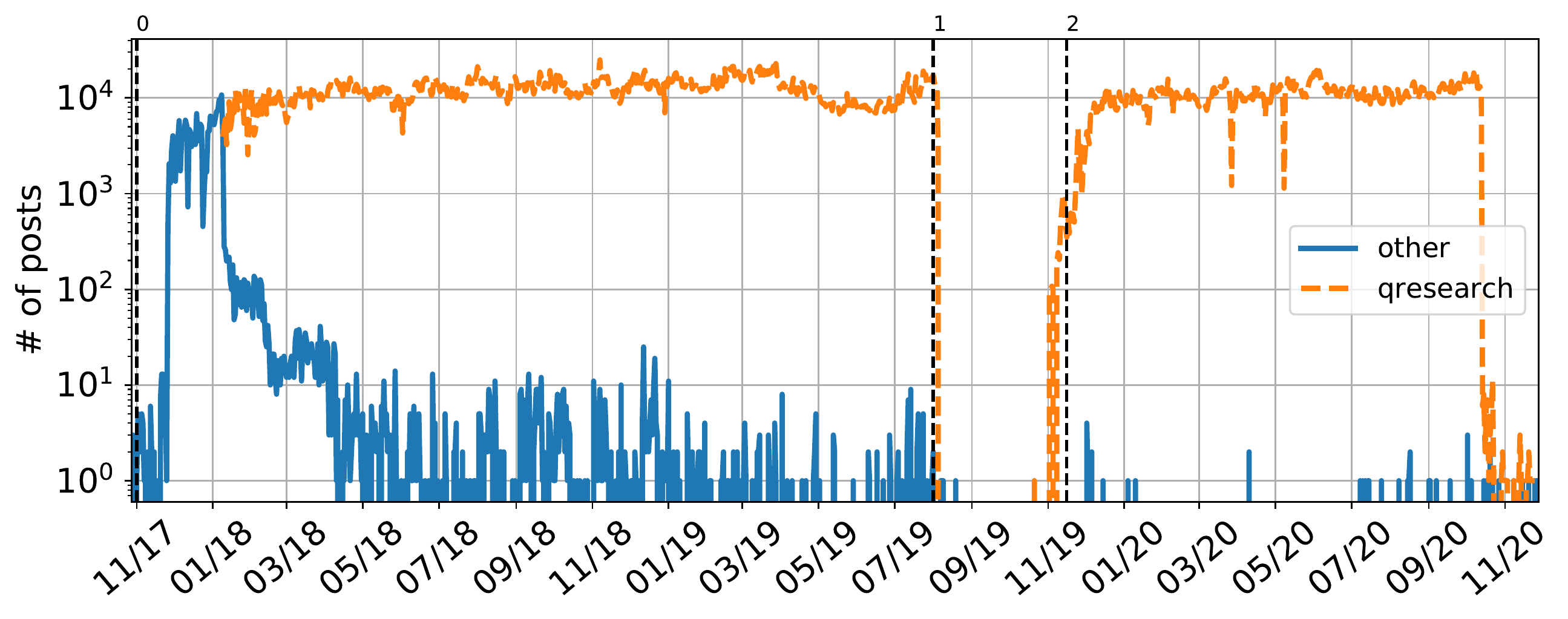}
  \caption{Number of daily posts on \dspol threads Q posted in plus all daily posts from 8kun's QAnon-related boards (labeled as ``other''), as well as all daily posts from \qresearch.}
  \label{fig:qresearch_posts}
\end{figure}

\descr{4chan and 8kun.}
We then measure Q-related activity on the imageboards between October 2017 and December 2020.
Several major events occurred on them which impacted the conspiracy, e.g., 8chan shutting down and resurfacing as 8kun.
We outline the events impacting the activity in Table~\ref{tab:events} and use these to annotate the longitudinal activity of QAnon imageboards in Figure~\ref{fig:qresearch_posts}.
Posting activity on \qresearch grew since November 2017, reaching above 10,000 posts per day in January 2018.
After, \qresearch almost always saw 10K posts per day, except for some days in June and December 2018, and May and June 2019.
Note that \qresearch was the main QAnon discussion community with orders of magnitude more activity than all the other boards combined.

\begin{figure}[t]
  \centering
  \includegraphics[width=0.75\columnwidth]{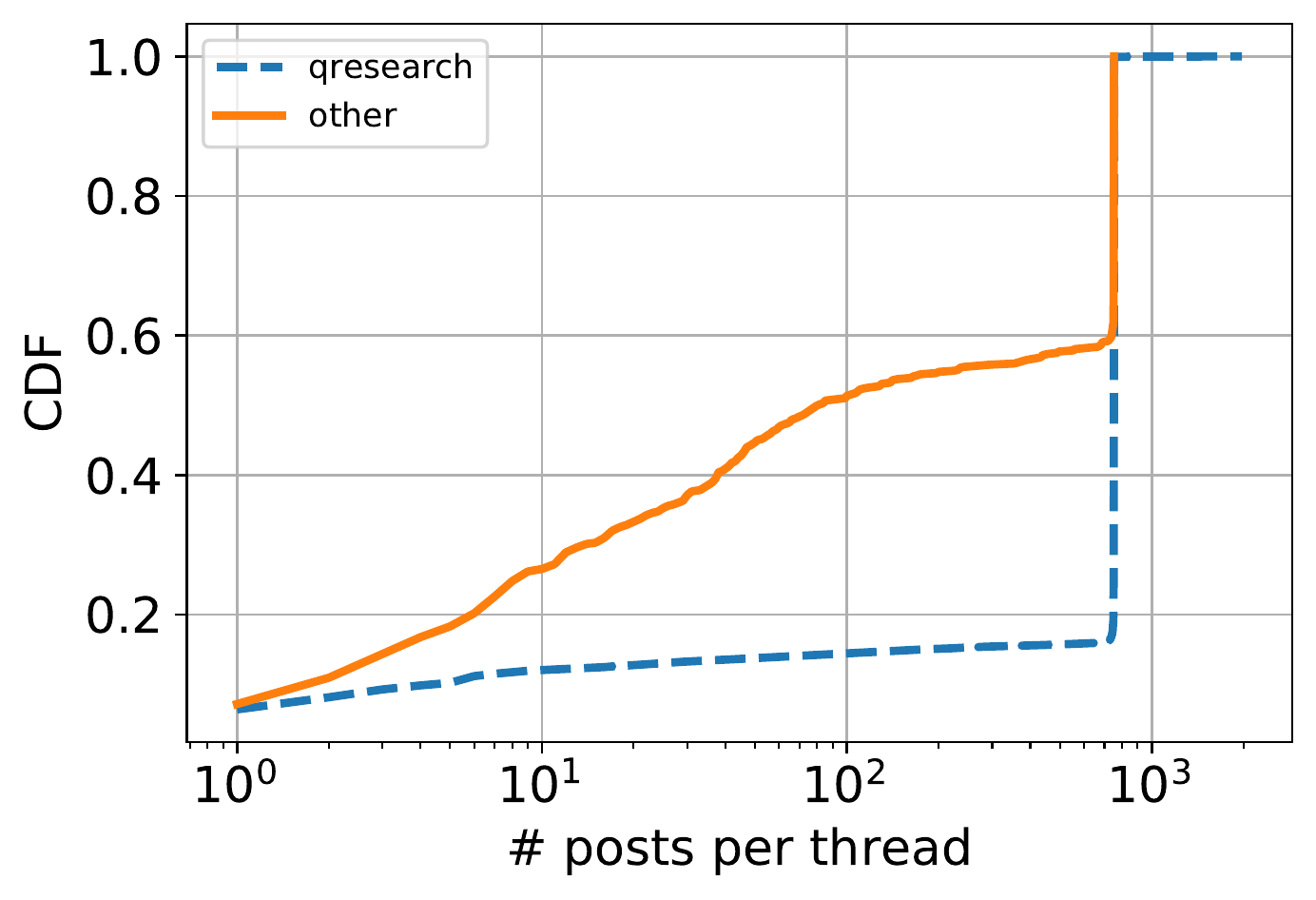}
  \caption{CDF of \#posts per thread on qresearch and other boards.}
  \label{fig:posts_per_thread_cdf}
\end{figure}

We also calculate the number of posts per thread across all \qresearch threads in Figure~\ref{fig:posts_per_thread_cdf}.
This shows that not only are \qresearch threads the largest, but they are also significantly larger than \dspol ones~\cite{papasavva2020raiders}.

\begin{table}[t]
  \centering
  \small
  \setlength{\tabcolsep}{2.5pt}
  \begin{tabular}{@{}rp{6cm}l@{}}
  \toprule
  {\textbf{\begin{tabular}[t]{@{}c@{}}Event \\ ID\end{tabular}}} & \multicolumn{1}{c}{\textbf{Description}}                                                            & \multicolumn{1}{c}{\textbf{Date}} \\ \midrule
  0                                                                                & Q started posting on 8chan~\cite{paris2017}.                              & 2017-11                        \\
  1                                                                                & 8chan went offline~\cite{8chandown}.  & 2019-08                      \\
  2                                                                                & 8kun replaced 8chan~\cite{Siegel2019}. & 2019-11                        \\
  \bottomrule
  \end{tabular}%
  \caption{Events depicted in Figure~\ref{fig:qresearch_posts}.}
  \label{tab:events}
  \end{table}
  
\subsection{Take Aways}\label{sec:54}
Overall, we find that Q discusses, among other things, governments controlled by despots and the duty of the people to revolt against it, often using the same language as the US Founding Fathers, e.g., excerpts from the US Declaration of Independence %
appears in ten Q drops:
\emph{
\begin{quotation}
\noindent But when a long train of abuses and usurpations, pursuing invariably the same Object evinces a design to reduce them under absolute Despotism, it is their right, it is their duty, to throw off such Government, and to provide new Guards for their future security.
\end{quotation}}

\noindent Also, Q drops are incredibly cryptic and incoherent.
Manual inspection of the Q drops sheds further light on the Perspective scores; we find that Q does not tend to include threatening content in the drops and that these are often extremely incoherent, consisting of short sentences, definitions, and various excerpts from movies and official documents.

This highlights how it is not the Q drops to be openly toxic/threatening or calling for violence, but rather the interpretations of the communities built around the conspiracy and the actors with vested interests that weaponize it.
We also demonstrate that bans on one platform or community are not enough to stop the spread of the conspiracy across platforms; however, it did make a significant impact in stopping the spread of QAnon aggregation site links on Reddit.
That is, banning main actors or communities involved in QAnon-related content did reduce the spread on the platform. 
In other words, this suggests that platforms will have to make a coordinated effort to reduce the spread of conspiratory content.
After Reddit banned the largest Q-focused subreddits, the majority of other Q-focused subreddits essentially died out.
However, the sharing of Q drops continued and spread to other subreddits at a minimal degree, and significantly less than before, hinting that actors may have taken content to other platforms instead.

Across Reddit and Voat, activity is driven by a small number of accounts. 
However, even the accounts which posted less remained active and participated in the discussion.

\section{Discussion \& Conclusion}\label{sec:conclusion}
This work presented a data-driven, multi-platform, multi-axes analysis of the QAnon conspiracy theory.
Our study of Q drops from six aggregation sites yielded several findings.
First, there are meaningful discrepancies in what is considered a canonical drop by different aggregation sites.
Next, we analyzed the content of 4,949 canonical Q drops, finding clear topics related to religion and calls to revolutionary action to defend freedom.
We also found statistically significant stylometric differences in Q drops per tripcode, indicating that a single person did not author them.
Finally, we showed how even though Reddit banned the two QAnon subreddits credited with helping the conspiracy theory go mainstream, links to Q drop aggregation sites still appear, primarily in right-wing-oriented subreddits.

\descr{Study Implications.}
There are several implications of our results; not just for understanding QAnon, but also for studying conspiracy theories.
There are relatively few Q drops, and the majority of activity ends up involving the dissemination and interpretation of them.
We also show that social networks cannot rely on open-source models, such as Google's Perspective API, to effectively detect conspiratory content, as it is not classified as toxic or threatening.

Moreover, we show that adherents often discuss and follow the conspiracy's updates but do not contribute towards dissemination.
This suggests that the conspiracy itself is coordinated by a handful of users.
Finally, while Reddit did take ``locally'' successful efforts to curb the spread of QAnon, Q drops were still discussed via links to aggregation sites.
In fact, the aggregation sites themselves, although not in perfect agreement, offer an additional degree of resilience to mitigation strategies like deplatforming.

\descr{Does the Q persona matter?}
Q is somewhat unique in that they are anonymous (and only active on anonymous and ephemeral platforms).
Our findings suggest that it is unlikely Q is a single individual, %
as we add to the growing body of evidence pointing towards a deliberate takeover of the Q persona, and coordination with actors intending to influence the 2020 US Presidential Elections~\cite{bbcQelections}.
This raises questions about how QAnon and future online conspiracies will evolve from a higher level.
While it might appear illogical for people to believe some of the basic tenets of this conspiracy, it is evident that adherents are motivated by actor(s) originating on a fringe social media platform with a known history of trolling~\cite{phillips2015we}.

Furthermore, tripcodes, the only 4chan and 8kun authentication mechanism, have changed multiple times as QAnon evolved and were subsequently ``compromised.''
(Aggregation sites even omitted Q drops related to Q's fighting over originality.)
Since Q is supposedly a government official, why would they choose imageboards to share classified information, despite the known insecurity of tripcodes?
Yet, given the importance of the Q persona, and the impact of the related conspiracy, tripcode compromise would significantly impact the content adherents would be presented with. %
In fact, adherents were motivated by the interpretation of the writings of what is likely to have started as a troll and reasonably controlled by more than one entity over time.
Moreover, when 8chan went down in August 2019, Q stayed silent and did not share updates to adherents via other means or platform~\cite{insiderQanalysis}. 
Why did Q fail to do so, albeit the ongoing crusade against the evil cabal is so crucial to the US democracy?

These considerations give a whole different ``meaning'' to the involvement of the QAnon followers in one of the most brazen and shocking attacks on US democracy%
~\cite{capitolriot2}.
Overall, QAnon originated on a niche platform like 4chan, moved to an even \emph{more} niche platform like 8chan, yet had and still has a \emph{global} impact.

\descr{Directions for future work.}
The adoption of drop aggregation sites raises questions about how effective deplatforming actions might be in preventing susceptible individuals from becoming adherents.
Further, these aggregation sites hold substantial power; since drops originally appear on anonymous and ephemeral imageboards, aggregation sites ultimately control what information adherents are left to interpret.
With this in mind, we believe there is a fruitful line of work that can build on our study; in particular, we encourage analysis on how drops are interpreted, both from the perspective of what they ``mean'' to adherents and from the perspective of which set of drops have the most impact on the theory's evolution.

Finally, %
we plan to dive deeper into understanding how Q drops are interpreted and how this changes over time.
While beyond the scope of this paper, QAnon provides a rich body of source material to develop a theoretical framework that explains just how the explicit lack of authentication of authoritative sources is exploited when considered from a socio-technical perspective.
Last but not least, given that aggregation sites are the ``bible'' of the conspiracy, we encourage future work that analyzes how \emph{exactly} these drops are discussed on even more platforms, like 
Parler~\cite{aliapoulios2021parler} and Twitter~\cite{mcquillan2020cultural}.

\descr{Acknowledgments.}
This work was partially funded by the National Science Foundation under Grant No. 1942610, 2114407, 2114411, and 2046590, the UK EPSRC grant EP/S022503/1 that supports the UCL Center for Doctoral Training in Cybersecurity, as th UK's National Research Centre on Privacy, Harm Reduction, and Adversarial Influence Online (REPHRAIN, UKRI grant: EP/V011189/1).
Any opinions, findings, and conclusions or recommendations expressed in this work are those of the authors and do not necessarily reflect the views of the funders.

\small
\bibliographystyle{abbrv}
\bibliography{references}

\end{document}